\documentclass[12pt,preprint]{aastex}
\usepackage{natbib}
\usepackage{multirow,color}
\usepackage{bbding}
\usepackage{amssymb}
\usepackage{ulem}
\usepackage{color}
\usepackage{longtable}
\usepackage{url}
\usepackage[displaymath]{lineno}

\def\gsim{\;\lower4pt\hbox{${\buildrel\displaystyle >\over\sim}$}\;}
\def\lsim{\;\lower4pt\hbox{${\buildrel\displaystyle <\over\sim}$}\;}
\def\grls{\;\lower4pt\hbox{${\buildrel\displaystyle >\over <}$}\;}

\newcommand{\jkt}{}

\begin{document}

\title{The causes of quasi-homologous CMEs} 

\author{Lijuan Liu \altaffilmark{1,2,3}, 
Yuming Wang \altaffilmark{1,4}, 
Rui Liu \altaffilmark{1,3}, Zhenjun Zhou \altaffilmark{1,5}, M. Temmer \altaffilmark{6}, J. K. Thalmann \altaffilmark{6}, Jiajia Liu \altaffilmark{1,4}, Kai Liu \altaffilmark{1,4}, Chenglong Shen \altaffilmark{1,4}, Quanhao Zhang \altaffilmark{1,5}, A. M. Veronig \altaffilmark{6}}
\email{ymwang@ustc.edu.cn}
\email{ljliu@mail.ustc.edu.cn }


\altaffiltext{1}{CAS Key Laboratory of Geospace Environment,
Department of Geophysics and Planetary Sciences, University of Science and
Technology of China, Hefei, Anhui, 230026, China}
\altaffiltext{2}{School of Atmospheric Sciences, Sun Yat-sen University, Zhuhai, Guangdong, 519000, China }
\altaffiltext{3}{Collaborative Innovation Center of Astronautical Science and Technology, China}
\altaffiltext{4}{Synergetic Innovation Center of Quantum Information \& Quantum Physics, University of Science and Technology of China, Hefei, Anhui 230026, China}

\altaffiltext{5}{Mengcheng National Geophysical Observatory, University of Science and Technology of China}
\altaffiltext{6}{Institute of Physics/IGAM, University of Graz, Universit\"atsplatz 5/II, 8010 Graz, Austria}

\begin{abstract}
{In this paper, we identified the magnetic source locations of $142$ quasi-homologous (QH) coronal mass ejections (CMEs), of which $121$ are from solar
cycle (SC) $23$ and $21$ from SC $24$. Among those CMEs, $ 63\%$ originated from the same source location as their predecessor (defined as S-type), while $37\%$ originated from a different location within the same active region as their predecessor (defined as D-type). Their distinctly different waiting time distribution, peaking around $7.5$ and $1.5$ hours for S- and D-type
CMEs, 
suggests that they might involve different physical mechanisms with different characteristic time scales. 
Through detailed analysis based on non-linear force free (NLFF) coronal magnetic field modeling
of two exemplary cases, we propose that the 
S-type QH CMES might involve a recurring energy release process from the same source location (by magnetic free energy replenishment), 
whereas the D-type QH CMEs can happen when a flux tube system 
disturbed by a nearby CME.
}

\end{abstract}

\section{Introduction} \label{sec:intro}

Coronal mass ejections, huge expulsions of plasma and magnetic fields from the solar corona, are among the drivers of hazardous space weather. 
Besides the knowledge on the propagation of a CME in interplanetary space, a successful space weather forecast also requires a precise understanding of the physical mechanisms behind CMEs, as well as their relation to other phenomena in the solar atmosphere.  
CMEs may originate from either active regions (ARs) 
or quiescent filament regions \citep[e.g.,][]{Schmieder_2006, Webb_2012}. Statistical studies suggest 
that about two thirds of CMEs originate from ARs, although  
the percentages vary from 
$63\%$ to $85\%$ 
in different studied samples \citep{SUBRAMANIAN_2001, Zhougp_2003, Chen_Wang_2011}.   
The flare and CME productivity of different ARs varies \citep[e.g.,][]{Tian_2002, Akiyama_2007, Chen_Wang_2011, Lliu_2016}. Some ARs barely produce an eruption, some produce numerous subsequent flares without accompanying CME \citep[e.g.,][]{Thalmann_etal_2015, Sun_etal_2015, Lliu_2016}, and some others can generate many flare-associated CMEs within a short duration. 
 It appears that ARs which accumulate large amounts of magnetic free energy tend to produce a larger number and more powerful flares and CMEs than ARs with a small magnetic free energy budget \citep[e.g.,][]{2010ApJ...713..440J,2014ApJ...788..150S}. Additionally, the larger a flare, the more likely it is accompanied by a CME \citep[e.g.,][]{2008ApJ...673.1174Y}. The triggering mechanism of a CME itself, however, is most likely determined by the involved magnetic field topology, both, of the unstable CME structure and its AR environment.
 
CMEs are termed ``homologous'' when they originate from the same region within an AR and exhibit a close morphological resemblance in coronal and coronagraphic observation \citep{Zhang_Wang_2002, Chertok_2004, Kienreich_2011, Li_2013}. 
However, CMEs may originate from different parts of an AR, and/or even have different appearances. 
Following \cite{Wang_2013}, we use the term ``quasi-homologous" CMEs, to denote subsequent CMEs that originate from 
the same AR, but disregarding their detailed magnetic source locations and appearances.

Statistical analysis of the waiting times of QH CMEs has been performed by  \cite{Chen_Wang_2011} and \cite{Wang_2013} in order to explore the physical nature of their initiation.  
The waiting time is defined as the time interval between the first appearance of a CME and that of its immediate predecessor in coronagraphic images.
The waiting time distribution for QH CMEs observed during 
$1997-1998$ consists of two components separated by 15~hours, where only the first component 
clearly exhibits the shape of a Gaussian, peaking around 8~hours \citep{Chen_Wang_2011}. This is significantly different from the 
waiting times of CMEs in general, appearing in the form of a Poisson distribution \citep{Moon_2003}. 
When only considering the QH CMEs that originated from the super ARs in solar cycle 23, the 
separation between the two components increases to about 18~hours, while the peak of the first component shifts to 7~hours 
\citep{Wang_2013}. 
CMEs with waiting times less than 18~hours, i.e.~the ones which contribute to the Gaussian component, are thought to have a close physical connection.

In addition, numerical simulations reveal that successive eruptions from a single AR may be driven by continuous shearing motions on the photosphere, the emergence of twisted magnetic flux tubes, reconnection between emerging and pre-existing flux systems, or perturbations induced by a preceding eruption \citep[e.g.,][]{Devore_2008, Mactaggart_2009, Soenen_2009, Torok_2011, Chatterjee_2013}.


Most CME-productive ARs exhibit a complex photospheric magnetic field configuration, consisting of a mix of {\jkt flux concentrations. 
Adjacent flux concentrations with opposite polarities, which may hold a flux tube, are separated by 
a polarity inversion line (PIL). 
Depending on the polarity pairs  
}
being present within an AR, a number of PILs (of different length and shape) may be present. Note that in some conditions, more than one {\jkt polarity pairs}  
are closely located in the vicinity of each other, with same polarity placed at the same side, forming a long PIL; i.e., a long PIL may be spanned by more than one flux tubes {\jkt , thus, be divided into different parts}.
Based on this, \citet{Chen_Wang_2011} envisaged three possible scenarios for QH CMEs to occur: successive CMEs may originate (i) from exactly the same part of a PIL, (ii) from different parts of the same PIL, (iii) from different PILs within the same AR.  The first scenario has been envisaged as the recurring release of quickly replenished magnetic energy/helicity. The other two have been regarded as scenarios where neighbouring flux tubes, either spanning different parts of a common long PIL or spanning distinctly different PILs, are disturbed, become unstable and erupt. Since the peak value of the waiting time distribution may represent the characteristic time scale of the most probable involved physical  process (either recurring release of the magnetic free energy or destabilization), we further explore the database of \citet{Wang_2013} in this work, in order to depict the most probable scenarios for QH CMEs to occur. 

\section{Identification and classification of QH CMEs}\label{sec:source}

\subsection{Event sample} \label{subsec:smaple}

The event sample of \citet{Wang_2013} consists of 281 QH CMEs that originated from 28 super ARs in SC~23. 
The CMEs are all listed in the {\it SOHO}/LASCO CME catalog\footnote{\url{http://cdaw.gsfc.nasa.gov/CME_list/}} \citep{2004JGRA..109.7105Y},
and their source ARs have been determined\footnote{\url{http://space.ustc.edu.cn/dreams/quasi-homologous_cmes/}} following the process described in \citet{wang_2011_st}. 
It is based on a combination of flares and EUV dimmings or waves, as they are strong evidence for the presence of CMEs. 
In particular, in the present work, we use localized flare-associated features, such as flare kernels, flare ribbons, and post-flare loops in order to determine the (portions of the) PIL relevant to the individual CME. 

Another two well-studied CME-rich ARs, 
NOAA AR~11158 and~11429, are added into the sample for detailed case study, as they were observed during the SDO \citep{Pesnell_2012} era, allowing an in-depth study of the associated flare emission using coronal imagery from AIA \citep{Lemen_etal_2012} and the involved coronal magnetic field structure and evolution based on vector magnetic field measurements from HMI \citep{Schou_2012,Hoeksema_etal_2014}.
Out of all of the events, $188$ QH CMEs exhibit a waiting time of less than $18$ hours, thus we assume them to be physically connected. 

Due to limitations in the observational data, not all of the 188 QH CME events could be successfully assigned to one of the three categories introduced  
above, i.e., whether to originate, from the exactly same portion of a PIL, 
from different portions of the same PIL, or a different PIL 
within the same AR as their predecessor. 
The CME assigned to the first category (the latter two categories) are defined as S-type (D-type) QH CMEs. 
Note that QH CMEs were assigned to the second category, only when they originated from totally different portions of a long PIL (with non-overlapped post-flare loops, ribbons, etc.). 
In total, we were able to clearly identify the magnetic sources of $142$ 
QH CMEs.  Among them, $90$ are classified as S-type, 
accounting for $ 63\% $; $52$ are classified as D-type, accounting for $37\%$. 
Selected QH CMEs are discussed in detail in the following two subsections, in order to demonstrate the identification process. The preceding CME is referred to as CME$1$, and the following CME is referred to as CME$2$. The associated flares are accordingly referred to as flare$1$ and flare$2$.

\subsection{Examples of S-type QH CMEs} \label{subsec:s_example}

{\flushleft\textbf{S-type QH CME from AR NOAA~9026}}

AR NOAA~9026, observed in the form of a large bipolar sunspot region with a $\delta$-spot, (Fig.~\ref{fig:eru_9026}(a)), was a highly CME-productive AR that launched at least $12$ CMEs during its disk passage.
Note that the strong positive polarity at the $[-300\arcsec,320\arcsec]$ in Fig.~\ref{fig:eru_9026}(a) belongs to AR 9030. 
Fig.~\ref{fig:eru_9026} shows the magnetic source location, morphology and the time evolution of an S-type CME and its predecessor that both originated from the main PIL, 
located within the yellow box L$1$ 
in Fig.~\ref{fig:eru_9026}(a). 
Fig.~\ref{fig:eru_9026}(b) - (d) show the evolution of the CME1-associated M$7.1$ flare$1$, 
as observed by {\it TRACE} \citep{1999SoPh..187..229H} at 1600~\AA,
while the white-light appearance of CME1 in LASCO/C2 \citep{Brueckner_1995} 
is shown in Fig.~\ref{fig:eru_9026}(e). 
Fig.~\ref{fig:eru_9026}(f) - (i) show the corresponding features of CME2 and its associated X$2.3$ flare$2$. From Fig.~\ref{fig:eru_9026} it is evident
that the chromospheric ribbons of both, flare$1$ and flare$2$, appear and evolve along the same part of the main PIL of the AR. Thus, CME$2$, with a waiting time of one hour, is classified as an S-type CME.

{\flushleft\textbf{S-type QH CME from AR NOAA~9236}}

AR NOAA~9236 
produced more than 15 CMEs during its disk passage. The AR hosted a $\delta$-spot of positive polarity surrounded by scattered elements of 
negative polarity (see Fig.~\ref{fig:eru_9236}(a)). 
The PIL of interest is located within the yellow box L$1$. 
The {\jkt two CMEs} (see Fig.~\ref{fig:eru_9236}(e) and \ref{fig:eru_9236}(i)) were 
associated with an X$2.3$ and an X$1.8$ flare, respectively.
The according {\it TRACE} 1600~\AA\ observations  
(Fig.~\ref{fig:eru_9236}(b) - (d) and Fig~\ref{fig:eru_9236}(f) - (h), respectively) reveal that 
the ribbons of the two 
flares appeared at the same location. CME2 had a waiting time of 7 hours and is thus classified as an S-type event. Note that these two CMEs were also classified as homologous events in \citet{Zhang_Wang_2002} and \citet{Chertok_2004}. 

{\flushleft\textbf{S-type QH CME from AR NOAA~11158}}

AR NOAA~11158 was the first super AR in SC 24 and produced more than 10 CMEs during disk passage. 
A pair of opposite polarities in the quadrupolar AR (outlined by the yellow box L$1$ in Fig.~\ref{fig:eru_11158}(a)) produced a number of CMEs within two days. 
Most of the CMEs were front-side, narrow events and missed by {\it LASCO}. However, they were all well captured by {\it STEREO}/COR1 \citep{kaiser_etal_2008}. 
The pair of CMEs shown in Fig.~\ref{fig:eru_11158}(e) and \ref{fig:eru_11158}(i) were associated with an M2.2 and a C6.6 flare, respectively (see Fig.~\ref{fig:eru_11158}(b) - (d) and Fig.～\ref{fig:eru_11158}(f) - (h)). 
The mass ejections (marked by the white arrows in Fig.~\ref{fig:eru_11158}(d) and (h)) shared the same source location. CME$2$, with a waiting time of 2.2 hours, is thus classified as an S-type QH CME.  
The cyan curve A1 in Fig.~\ref{fig:eru_11158}(a) indicates the projection of the flux rope axis along the 
related PIL at Time1, i.e., before the occurrence of CME1. The pink curve A2 indicates the flux rope axis position at Time2, i.e., at a time instance after CME1 happend but before CME2 was launched. The lines C1 and C2  mark the position of two vertical cuts that will be used to derive some flux rope parameters at the two time instances. For details see Sec. \ref{cases:11158}.

\subsection{Examples of D-type QH CMEs} \label{subsec:d_example}

{\flushleft\textbf{D-type QH CME from AR NOAA~10030}}

AR NOAA~10030 adhered to a quadrupolar 
configuration (see Fig.~\ref{fig:eru_10030}(a)) and produced at least 8 CMEs during disk passage. A CME and its QH predecessor are shown in Fig.~\ref{fig:eru_10030}(i) and \ref{fig:eru_10030}(e). 
The yellow boxes L1 and L2 in Fig.~\ref{fig:eru_10030}(a) enclose the pairs 
of opposite polarities, relevant to the respective CMEs, CME1 and CME2, and defining the accordingly relevant PILs (PIL1 and PIL2, respectively).
CME$1$ was accompanied by a X3.0 flare 
(see Fig.~\ref{fig:eru_10030}(b) - (d)). 
Though an extra ribbon appeared in the positive polarity in L$2$ in Fig.~\ref{fig:eru_10030}(b), 
the helical structure marked by the white arrow in Fig.~\ref{fig:eru_10030}(b), and the observed chromospheric ribbons 
support that CME$1$ originated from L1. 
Fig.~\ref{fig:eru_10030}(f) - (h) show the time evolution of the chromospheric ribbons of the 
CME2-associated M1.8 flare$2$, clearly aligned with PIL$2$. 
CME$2$, with a waiting time of $1$ hour, thus is classified as a D-type CME. 
Already \citet{Gary_2004} demonstrated that the two CMEs should have originated from two different magnetic flux tube systems, and further argued that the observational signatures matched  
a breakout scenario.  

{\flushleft\textbf{D-type QH CME from AR NOAA~10696}}

AR NOAA~10696, similar to NOAA~9236, 
consisted of a concentrated negative polarity region surrounded by scattered small positive polarity patches (see Fig.~\ref{fig:eru_10696}(a)). It produced more than 12 CMEs. The yellow boxes L$1$ and L$2$ in Fig.~\ref{fig:eru_10696}(a) mark the 
source locations of CME1 and CME2, respectively.  
Fig.~\ref{fig:eru_10696}(b) - (d) and \ref{fig:eru_10696}(f) - (h) show the evolution of the associated M5.0 and M1.0 flare, respectively. Fig.~\ref{fig:eru_10696}(e) and \ref{fig:eru_10696}(i) show the appearance of the CMEs in {\it LASCO}/C2. The white arrows in Fig.~\ref{fig:eru_10696}(d) and \ref{fig:eru_10696}(h) mark the post-flare loops associated with the two CMEs, 
further supporting that they originated from different flux tube  systems. CME$2$ had a waiting time of 2.8~hours and is therefore classified as a D-type QH CME.

{\flushleft\textbf{D-type QH CME from AR NOAA~11429}}

AR NOAA~11429, a super active AR in SC 24, produced more than 12 CMEs during disk passage. The AR exhibited a complicated topology with a $\delta$-spot.
The two yellow boxes L$2$ and L$1$ in Fig.~\ref{fig:eru_11429}(a) mark the magnetic source locations of a CME and its QH predecessor. The cyan curve A$1$ indicates the projection of the flux rope axis along PIL$2$ at Time$1$, i.e., before the occurrence of CME1. The cyan line C$1$ mark the position of a vertical plane that perpendicular to A1 at Time1. The pink curves A$2$ and C$2$ are corresponding axis and plane for PIL$2$ at Time$2$, i.e., a time instance after CME1 happened but before CME2 was launched. See more details in Sec.~\ref{cases:11429}. 
The time evolution of the flares that accompanied the two CMEs, an X5.4 and an X1.3 flare, 
is shown in Fig.~\ref{fig:eru_11429}(b) - (d) and \ref{fig:eru_11429}(f) - (h), respectively. 
The white arrow 
in Fig.~\ref{fig:eru_11429}(h) marks the post-flare loops of 
CME$2$, while the black arrows in Fig.~\ref{fig:eru_11429}(f) - (h) mark the post-flare loops of CME$1$. 
CME$2$, with a waiting time of one hour, is classified as a D-type CME, in agreement with its classification by \citet{Chintzoglou_2015}.

\subsection{Waiting-time Distribution} \label{subsec:wt}

The waiting time distribution of the $188$ CMEs (with waiting times $<$ $18$~hours) is shown as a black curve in Fig.~\ref{fig:wt}, exhibiting a Gaussian-like distribution with a peak at about $7.5$~hours, suggesting that they are physically related.  
The distributions of  
precisely located S- and D-type QH CMEs, 
are shown as a blue curve and a red curve in Fig.~\ref{fig:wt}, respectively.
The two are distinctly different from each other: the former peaks at $7.5$ hours while the latter peaks at $1.5$ hours,  
strongly supporting that these two types of QH CMEs may be involved into different physical mechanisms. 
 Another slightly lower peak appears around 9.5 hours in the waiting time distribution of D-type QH CMEs.
One possible reason is that in some cases, a CME triggers a D-type QH CME in a short interval of around 1.5 hours, after which the first CME's source region undergoes a energy replenishment and produces another QH CME with a interval around 7.5 hours. However, the third CME would be {\jkt classified} as a D-type, as it originates from a different source location from its predecessor, with a waiting time of around 6 hours. Considering the 3 hours bin size of the distribution, a peak around 9 hours may be reasonable. Another possible reason is that those D-type QH CMEs with waiting times around 9.5 hours may follow a different mechanism from the ones with short waiting times (around 1.5 hours). The work aims to find the most possible (but not only) scenario for the two types of QH CMEs.

In order to explore the different underlying mechanisms, 
the aforementioned S-type CME in AR 11158 and D-type CME in AR 11429 are analyzed in details in the next section. 
These two cases  
were observed during the {\it SDO} era, allowing for sophisticated modeling of the three dimensional (3D) coronal magnetic field, based on the measurements of the photospheric magnetic field vector at a high spatial resolution from {\it SDO}/HMI.

\section{Coronal magnetic field topology of S- and D-type CMEs}\label{sec:cases}

\subsection{Method}\label{subsec:method} 
It is widely accepted that the expulsion of a CME is determined by the inner driving force (associated to, e.g., an erupting flux rope) and the external confining force (exerted by the large-scale, surrounding coronal magnetic field) \citep[e.g.,][]{Wang_zhang_2007, Liuy_2008, Schrijver_2009}. 
In order to investigate the involved mechanisms, the knowledge of the 3D coronal magnetic field 
is {\jkt necessary}. 
A method developed by Wiegelmann \citep{Wiegelmann_2004, Wiegelmann_2012} is employed to 
the two selected cases, to reconstruct the 3D potential (current-free) and nonlinear force-free (NLFF) fields in the corona, based on the surface magnetic vector field measurements from HMI.
  
{A magnetic flux rope, characterized by magnetic fields twisted about a common axis, may become unstable and act as a driver for an eruption \citep[e.g.,][]{Tamari_etal_1999, Torok_Kliem_2005}. A flux rope can be identified 
using a combination of topological measures deduced from the employed NLFF models, e.g.,  in the form of the twist number $T_w$  and the squashing factor $Q$ \citep{rliu_2016}.
$T_w $ gives the number of turns by which two infinitely approaching field lines, i.e., two neighbouring field lines whose separation could be arbitrarily small, wind around each other, and is computed by 
\begin{equation}\label{eq:twi}
T_w = \frac{1}{4\pi}\int_L\alpha dl
\end{equation}
where $\alpha$ is the force-free parameter, $dl$ is the length increment along a magnetic field line, $L$ is the length of the field line \citep{MBerger_2006, rliu_2016}. $Q$ is a measure of the local gradients in magnetic connectivity; regions with high values of $Q$ are referred to as Quasi-separatrix Layers (QSLs) \citep{Titov_2002, Titov_2007}.

The cross section of a flux rope 
with twisted field lines treading the plane, would be visible as 
a region of strong $T_w$ enclosed by a surface of high $Q$ values 
that separating 
the magnetic fields of the flux rope from its magnetic environment.  The location of the local extremum $T_w$ in the cross section of a coherent flux rope is a reliable proxy of the location of its central axis.  
Additionally, 
a cross section perpendicular to the axis of the flux rope (e.g., the section at the apex point of the flux rope axis) would allow the axis run through the plane horizontally, so that 
the in-plane vector field will show a clear rotational pattern around the axis, which is represented by the point where $T_w$ is maximal.

The external confining force can be measured by the decay index
\begin{equation}\label{eq:dec}
n=-\frac{\rm d\ln B_{ex}(h)}{\rm d\ln h}
\end{equation}
where $h$ is the radial height from the solar surface, $B_{ex}$ is the horizontal component of the strapping potential field above the AR. Basically, $n$ measures the run of the strapping field's confinement with height. 
Theoretical works predict the onset of torus instability when $n$ is in the range of $[1.5,2.0]$ 
\citep{Kliem_2006}, while observations of eruptive prominences suggest a critical value $n\sim 1$ \citep{Filippov_2013, SuY_2015}. 
It is suggested 
that the former value is representative for the top of the flux rope axis, while the latter value is typical for the location of magnetic dips 
that hold the prominence material \citep{Zuccarello_2016}. Therefore, $n=[1,1.5]$ are used as critical decay index values for our analysis. 
Torus instability sets in once the axis of the flux rope reaches a height in the corona at which the strapping potential fields decrease fast enough \citep{Torok_Kliem_2005}, thus the vertical distribution of $n$, along the axis of the flux rope, will hint at its instability.

Since a physical relation is assumed to exist between the QH CMEs (CME$2$ and its predecessor, CME$1$), we may expect 
a change in the 
magnetic field configuration of the CME2's source location after CME$1$, 
detectable in the form of a change of the related parameters defined above ($T_w$, $Q$ and $n$). Therefore, we deduce these parameters from the NLFF models (for $T_w$ and $Q$) and potential models (for $n$) of the pre-CME1, and post-CME$1$ (i.e., pre-CME2) corona as follows:
\begin{enumerate}
\item Locate the axis of the flux rope using the method of  \citet{rliu_2016}, which calculates the twist maps in many vertical planes at first, and traces the field line running through the peak $T_w$ point at each map. All traced field lines should be coinciding with each other if a coherent flux rope is present. The line is then considered as to represent the flux rope axis.
\item Calculate $T_w$ and $Q$ in a vertical plane perpendicular to the flux rope axis. The
in-plane vector field, $\vec{B}_\parallel$, can provide additional evidence of the presence of a flux rope in the form of a clear rotational pattern, centered on the flux rope axis' position. 
\item Calculate the decay index $n$ in a vertical plane, aligned with the flux rope axis and extending from the flux rope axis upwards, as a function of height in the corona.
\end{enumerate}

Using the above introduced models and concepts, we investigate the pre-CME1 and post-CME1 (pre-CME2) 
coronal magnetic field configuration of the mentioned two cases in ARs NOAA~11158 (Sect.~\ref{cases:11158}) and 11429 (Sect.~\ref{cases:11429}) in great detail. The quality of all the NLFF extrapolation in this paper is shown in Appendix~\ref{apd:qual}.

\subsection{S-type QH CME from AR NOAA~11158} \label{cases:11158}

As demonstrated in Sec.~\ref{subsec:s_example}, the S-type CME and its predecessor originated from the same PIL within NOAA~11158. 
We study the magnetic parameters at the CMEs' source location (L1) 
at two time instances: once before CME1, at 2011-02-14T17:10:12~UT (Time1), and once after CME1 but before CME2 at 2011-02-14T18:10:12~UT (Time2).

At both times, we find a flux rope structure from the constructed corona field  
(see Fig.~\ref{fig:rope_11158}(g) and (h)). 
The magnetic properties of the pre- and post-CME1 flux rope in a vertical plane perpendicular to its axis are shown in Fig.~\ref{fig:rope_11158}(a) - (c) and \ref{fig:rope_11158}(d) - (f) (from left to right: $Q$, $T_w$, and $\vec{B}_\parallel$)}, respectively. The footprints of the vertical planes at the two times 
are marked as C1 and C2 in Fig.~\ref{fig:eru_11158}(a). 
Their vertical extensions are indicated by the yellow lines in Fig.~\ref{fig:rope_11158}(g) and (h). 
At Time$1$ (pre-CME1), a region of strong twist 
(Fig.~\ref{fig:rope_11158}(b)) is surrounded by a pronounced $Q$-surface 
(Fig.~\ref{fig:rope_11158}(a)). 
The diamond symbols in Fig.~\ref{fig:rope_11158}(a) - (c) 
mark the location where $T_w$ is strongest, at $T=-1.94$, and are assumed to represent the 3D location of the flux rope axis, at a coronal height of $h\gtrsim2$\,Mm.
The in-plane vector magnetic fields, $\vec{B}_\parallel$ (Fig.~\ref{fig:rope_11158}(c)), 
show a clear rotational pattern, centered around the 
flux rope axis, suggesting a left-handedness of the flux rope, since the blue arrows indicate the vector fields with the normal components going into the plane.
The field lines passing through the 
strong twisted region are shown in Fig.~\ref{fig:rope_11158}(g) in cyan, 
even adhering to a Bald Patch (BP) \citep[a set of field lines that graze the photosphere at the PIL; see, e.g.,][]{Titov_1993}. A representative field line in the BP is plotted as a white line, which is determined by the criteria introduced in \citet[Formula 32,][]{Titov_1999}. 
 
At Time2 (post-CME1), the highest value of twist in the vertical plane perpendicular to the flux rope axis is found as $T_w=-2.11$, marked by the diamond symbols in Fig.~\ref{fig:rope_11158} (d) - (f). Again, a region of strong twist (Fig.~\ref{fig:rope_11158}(e)) 
is surrounded by a pronounced $Q$-surface (Fig.~\ref{fig:rope_11158}(d)), but located lower in the model corona (height of the flux rope axis $h\lesssim2$\,Mm). The fields traced from the high-$T_w$ region are 
shown in Fig.~\ref{fig:rope_11158}(h) as pink curves. For comparison, the outline of the flux rope at Time$1$ is shown again as cyan curves. The more potential arcade fields (white lines) are traced at Time2 but from the coordinates of the top of the flux rope at Time$1$. 


The direct comparison between the pre- and post-CME1 model magnetic field configuration suggests that the upper part of the flux rope  
might erupt during CME$1$, while the lower-lying part of the flux rope seems to remain.
In order to check the conjecture,  
we further trace the field lines within the pre-CME1 corona from exactly the same starting locations used for tracing the post-CME1 flux rope (i.e., the high-$T_w$ region enclosed by the high-Q boundary at Time2; see Fig.~\ref{fig:rope_11158}(d) and ~\ref{fig:rope_11158}(e)). The traced pre-CME1 field configuration (red lines in Fig.~\ref{fig:rope_11158}(h)) clearly differs from the post-CME1 field structure (pink lines in Fig.~\ref{fig:rope_11158}(h)), {\jkt which may suggest two possibilities: (i) the flux rope totally erupted during CME1, after which a new one emerged, or reformed; (ii) the flux rope underwent a topology change that part (not simple the upper part) of it was expelled during CME1, while the other part was left, being responsible for CME2. See Appendix~\ref{apd:11158_cme2} for some details for CME2. } 

We also calculated the unsigned vertical magnetic flux from the strong $T_w$ region in the aforementioned planes. No strong twist region exists outside of the flux rope, thus, instead of doing a image-based flux rope recognition, we directly select the regions with $|T_w|\gtrsim 1.25$. $|T_w|=1.25$ is 
a threshold value for kink instability \citep{Hood_1981, Torok_2003}. The flux is calculated by 
\begin{equation}\label{eq:flux}
\Phi_{1.25} = \int_{A_{1.25}}|\vec{B}_\perp| dA
\end{equation}
in which $\vec{B}_\perp$ is the magnetic fields perpendicular to the vertical plane, $dA$ is the element area. The planes are perpendicular to the axes of the pre- and post-CME1 flux ropes, thus the vertical magnetic flux can represent the axial flux of the flux rope. The unsigned vertical magnetic flux (given at the the header of Fig.~\ref{fig:rope_11158}(a) and ~\ref{fig:rope_11158}(d)) decreased from $4.52\times 10^{19} $\,Mx at Time1 to $3.10\times 10^{19} $\,Mx at Time2, {\jkt which can be due to either ejection or simple redistribution of twisted field lines, since twist is not supposed to be conserved during the flux rope evolution. However, CME1 has been confirmed to be related with source location L1 based on observation, as discussed in Sec.~\ref{subsec:s_example}, the decrease here is more likely to support a twist release through eruption rather than redistribution.  

We can not make a definite conclusion on whether the flux rope at Time2 is a partial eruption remnant, or is a newly emerged/reformed one. However, the pre-CME1 flux rope has a BP, and the post-CME1 rope has some nearly-potential loops right above it. Thus, 
we prefer} a partial expulsion model 
\citep{Gibson_2006}, consisting of a coherent flux rope with a BP, to explain the eruption process: the field lines in the BP are not free to escape so that during the writhing and upward expansion of the ends of the field lines, a vertical current sheet may form, along which internal reconnection may occur and finally split the flux rope into two parts. The white arcades in Fig.~\ref{fig:rope_11158}(h) could be the post-eruption loops, which may also support that part of the flux rope erupted with CME$1$.  

Fig.~\ref{fig:decay_11158}(a) and (b) shows the distribution of the decay index $n$ as a function of height above the flux rope axis, for Time$1$ and Time$2$, respectively. The projections of the flux rope axis at the two times 
are indicated by the curves A$1$ (for Time1) and A$2$ (for Time2) in Fig.~\ref{fig:eru_11158}(a).
The solid lines in Fig.~\ref{fig:decay_11158} indicate the height where $n=1$ and $n=1.5$. 
It is evident that, for both time instances, the vertical run of $n$ varies strongly along the flux rope, with the $n=1.5$ level being located at a height above 48\,Mm at one end, and around 16\,Mm at the other end of the flux rope.  
The height where $n=1$ varies less dramatically along the flux rope, and is located at the height around 10\,Mm. 
 Comparison of the $n=1.5$ level at Time1 and Time2 (represented by the dotted and solid curves in Fig.~\ref{fig:decay_11158}(b), respectively) suggests that the critical height at the south-eastern end (x=0\,Mm in Fig~\ref{fig:decay_11158}) is lowered by about 8\,Mm.  
In the remaining part of the flux rope, no significant change was detected, which indicates that the external confining force was not lowered significantly by the first eruption. The critical height, both before and after CME1, were located relatively low in the solar atmosphere (e.g., $n = 1$ at $h\approx10$\,Mm), but still far above the height of the flux rope axis (red lines in Fig.~\ref{fig:decay_11158}(a) and (b)) located below 3\,Mm at both times. The maximal $n$ at the flux rope axis reaches $0.80$ at Time1, and  $0.44$ at Time2,  which are both lower than the critical $n=1.5$ for torus instability. The results argue against torus instability in triggering the two QH CMEs. 

\citet{Sun_2012a} studied the long-term evolution of AR NOAA~11158 and showed that the fast emergence and continuous shear of a bipolar photospheric magnetic field (L1 in Fig.~\ref{fig:rope_11158}) accumulated a large amount of magnetic free energy before the onset of a series of QH CMEs. They showed that the emerging fields reconnected with pre-existing fields, which finally led to the eruptions. Together with our analysis, their results hint at a multi-stage energy release process during which the magnetic free energy is released due to the successive eruptions from the same bipolar region (L1 in Fig.~\ref{fig:rope_11158}). Meanwhile, the energy was replenished through the shearing motion and ongoing flux emergence. We also calculate the magnetic free energy in the entire extrapolation volume at the two time instances (shown as $E_F$ in Fig.~\ref{fig:rope_11158} (b) and (e)) by 
\begin{equation}\label{eq:ene}
E_F=\int_V \frac{B^2_N}{8\pi}dV-\int_V \frac{B^2_P}{8\pi}dV
\end{equation}
$B_N$ is the NLFF filed, $B_P$ is the potential field and $dV$ is the element volume. $E_F$ shows a slight increase by $5\%$ from Time1 ($2.06 \times 10^{32} $\,erg) to Time2 ($2.17 \times 10^{32} $\,erg), which is against the expectation that the magnetic free energy would decrease after CME1, since CME1 should have taken part of the free energy during the multi-stage energy release process. The slight increase could be due to the small fraction of the big, fast evolving AR that the erupting bipolar system account for, and/or the fast accumulation of the magnetic free energy by 
flux emergence and shear motions. Besides, the free energy calculated from the model coronal field has an uncertainty of around $10\%$ 
\citep{Thalmann_2008a}, so that no definite conclusion on the loss of free energy during CME1 could be made here. 

\subsection{The D-type QH CMEs from AR NOAA~11429} \label{cases:11429}

As discussed in Sec.~\ref{subsec:d_example}, a D-type CME and its predecessor originated from two different locations within NOAA~11429, separated by a waiting time of just 1~hour. A physical relation is assumed to exist between the two QH CMEs, thus, a change at the source location of CME2 after CME1, is expected (see Sec.~\ref{subsec:method}).
Therefore, we study the magnetic parameters at the source location (L$2$) of CME$2$ at two time instances in the following. Once before CME1, at 2012-03-06 23:46:14~UT (Time1), and once after CME1 but before CME2 at 2012-03-07 00:58:14~UT (Time2).

Fig.~\ref{fig:rope_11429} shows the $T_w$, $Q$, in-plane vector fields ($\vec{B}_\parallel$) maps and the traced flux ropes for AR 11429. Through checking the $T_w$ and $Q$ maps in many vertical cuts across PIL2, we found {\jkt three possible flux ropes at Time$1$. 
The peak $T_w$ point resides in the middle structure, thus, we again identified the axis of the middle} rope with the peak $T_w$ point and then place a plane perpendicular to the flux rope axis. The plane's footprint is marked as C1 in Fig.~\ref{fig:eru_11429}(a) and its vertical extent is marked by the yellow vertical line in Fig.~\ref{fig:rope_11429}(g). Fig.~\ref{fig:rope_11429}(a) - (c) show the distribution of $Q$, $T_w$ and $\vec{B}_\parallel$ calculated in the plane. The axis of the {\jkt middle} flux rope, with a peak value $T_w=1.86$, is indicated by diamond symbols. The in-plane vector field, $\vec{B}_\parallel$, displays {\jkt three} clearly rotational patterns with opposite handedness{\jkt , alternately}. This supports that there were {\jkt three} flux ropes present along PIL2 at Time1. {\jkt A configuration with two vertically arranged flux ropes, 
i.e., a so-called double-decker flux rope, has been studied~\citep{liu_2012a, 2014ApJ...792..107K}. However, a similar configuration, with three flux ropes presented here, is barely reported to our knowledge, we name it a triple-decker flux rope, analogically.} The blue arrows indicate the vector magnetic fields with vertical component going into the plane, thus the upper one {\jkt and the lower one (FR$^3_2$ and FR$^1_2$ in Fig.~\ref{fig:rope_11429}) is left-handed} (i.e., the in-plane vector field exhibits a counter-clockwise sense of rotation), while the {\jkt middle one (FR$^2_2$) is right-handed}.  {\jkt The square and triangle symbols in Fig.~\ref{fig:rope_11429} (a), (b) mark the position of the axes of FR$^3_2$ and FR$^1_2$ , with local peak values $T_w=-1.82$ and $T_w=-1.49$, respective. The plane is not perpendicular to the axes of FR$^3_2$ and FR$^1_2$, positions of which are not well corresponding with the rotational centers of the ropes' in-plane fields, thus the symbols are not marked in Fig.~\ref{fig:rope_11429}(c).} Fig.~\ref{fig:rope_11429}(g) depicts the structure of the flux ropes, {\jkt FR$^3_2$ in blue, FR$^2_2$} in orange and {\jkt FR$^1_2$} in cyan. 
A longer, strongly twisted rope (marked as FR$_1$ in Fig.~\ref{fig:rope_11429}(g)--(h)), is aligned with PIL$1$, and is resulted in CME$1$. The white lines in Fig.~\ref{fig:rope_11429}(g) represent some nearly-potential arcades above the flux ropes. Note that the south-western end of FR$_1$ was located closely to the {\jkt triple-decker} flux rope along PIL$2$, and part of the arcade field was overlying both, the south-west end of the CME1-associated flux rope and the eastern part of the {\jkt tripple-decker} flux rope. Therefore, we may assume that the eruption of FR$_1$ easily affected the {\jkt triple-decker flux rope through various ways, e.g., by removing the common overlying arcades, disturbance, compressing the neighbouring fields through expansion of the post-eruption loop system below the erupted flux rope, even reconnecting with the neighbour fields during expansion.}

At Time$2$ (see Fig.~\ref{fig:rope_11429}(d)--(f)), the upper {\jkt two flux ropes along PIL$2$ evidently disappeared from the extrapolated domain, } 
while the lower one was now located higher, with a peak value $T_w=-1.81$ (indicated by triangles) located at $h\sim6$\,Mm. The whole structure also appears expanded compared to that at Time$1$. The in-plane vector field, $\vec{B}_\parallel$, exhibits a rotational pattern around the maximum value of $T_w$, which is evidence for the presence of a flux rope (Fig.~\ref{fig:rope_11429}(f)). The footprint of the vertical plane is marked as C$2$ in Fig.~\ref{fig:eru_11429}(a) and its vertical extent is marked as a yellow line in Fig.~\ref{fig:rope_11429}(h). Field lines traced from the strong $T_w$ region at Time$2$ are shown in pink in Fig.~\ref{fig:rope_11429}(h). For comparison, the flux ropes which was present at Time$1$ is shown as cyan lines. Comparison of {\jkt FR$_2^1$} at Time1 and Time2 reveals that it elevated and expanded, as well as gained internal twist. The vertical magnetic fluxes calculated by Equ.~\ref{eq:flux} from the strong 
$T_w$ region ($|T_w| \gtrsim 1.25$) {\jkt of the lowermost structure of the triple-decker flux rope ($h \lesssim 5$\,Mm at Time1 and $h \lesssim 8$\,Mm at Time2) ,} 
i.e., the representation of the axial magnetic flux of the lower flux rope (shown at the headers of Fig.~\ref{fig:rope_11429}(a) and (d)) indicate an increase by $2.48$ times (from $2.28\times 10^{19} $\,Mx at Time1 to $5.66\times 10^{19} $\,Mx at Time2), supporting the enhancement of the twist. 
{\jkt The upper two flux ropes, with opposite handedness, clearly disappeared from the system with almost no remnant left behind. A QSL exists between the two ropes (strong Q line at around 8.5\,Mm in Fig.~\ref{fig:rope_11429}(a)). Thus, we prefer annihilation due to local reconnection started from the QSL, rather than expulsion, to be account for the absence of them at Time2. Annihilation of the ropes would cause decrease of the local magnetic pressure, which is likely to allow FR$^1_2$ to rise, expand and finally erupt, giving rise to the faint CME$2$. 
}


Further support for this scenario is given by the evolution of the observed chromospheric ribbons as shown in Fig.~\ref{fig:rib_11429}. At the beginning of flare$1$, two ribbons, labeled R$_1^1$ and R$_1^2$ in Fig.~\ref{fig:rib_11429}(a), expand on both sides of PIL$1$. While R$_1^2$ grew southward in time (Fig.~\ref{fig:rib_11429}(b)), two more faint and small ribbons, R$_1^3$ and R$_1^4$, became visible along PIL$2$ (Fig.~\ref{fig:rib_11429}(c)).  Comparison to the flux ropes shown in Fig.~\ref{fig:rope_11429}(g) and (h), this pair of ribbons indicate the involvement of {\jkt FR$_2^2$ and FR$_2^3$ in the magnetic process. The two ribbons showed no clear sign of development that departed from, or along the PIL, which may be evidence for a local, small scale reconnection process. FR$_2^2$ and  FR$_2^3$ should have reconnected and annihilated during the first eruption. }
After flare1/CME1, the lower flux rope became unstable as well and erupted, giving rise to a further pair of flare ribbons, R$_2^1$ and R$_2^2$, at the beginning of flare$2$. 

{\jkt Note that there still existed a flux rope at PIL$1$ after CME$1$, though we cannot determine whether it's a remnant or a newly emerged/reformed one. A similar analysis is performed across PIL1. 
}See Appendix~\ref{apd:11429_cme1} for details. 
The magnetic free energy in the extrapolated pre- and post-CME1 corona volume (shown as $E_F$ in Fig.~\ref{fig:rope_11429} (g) and (h)) shows a decrease of $25\%$ (from $10.61 \times 10^{32} $\,erg at Time1 to $8.01 \times 10^{32} $\,erg at Time2), which is beyond the uncertainty ($10\%$), implying a clear energy release with CME1.   

Fig.~\ref{fig:decay_11429}(a) and (b) shows the distribution of the decay index $n$ as a function of height above the axis of the lower flux rope at PIL2, for Time$1$ and Time$2$, respectively. The projection of the flux rope axis at the two times is indicated by the curves A$1$ and A$2$ in Fig.~\ref{fig:eru_11429}(a). 
The solid curves mark the height where $n=1$ and $n=1.5$. The height at which $n=1.5$ varies 
between $h=30$\,Mm and 50\,Mm along the flux rope axis, while the height at which $n=1$ shows a similar trend but at lower heights (about $15$\,Mm lower). The dotted lines in Fig.~\ref{fig:decay_11429}(b) are critical heights at Time$1$ for comparison. The red lines indicate the height of the flux rope axis, that both are lower than 6\,Mm at the two time instances. No significant change is found, suggesting that CME$1$ may not significantly lower the constraining force of the overlying field. At both times, the predicted critical height for the onset of torus instability ($n=1.5$) is located much higher in the corona than the axis of the flux rope. Also the observation-based critical height (where $n = 1$) is located clearly above the flux rope. The maximal $n$ at the flux rope axis is $0.59$ at Time1, $0.53$ at Time2, respectively, both lower than the critical value $n=1.5$, also suggests that torus instability may not have been the direct trigger for the two CMEs. 


We conclude for the D-type CME and its predecessor from AR NOAA~11429, their magnetic source regions were located very close to each other, and bridged by the same large-scale potential field arcade. The first occurring CME1 (associated to the flux rope along PIL1) {\jkt destabilized the magnetic environment of the nearby flux tube system (above PIL2), leading to the reconnecting annihilation of the upper two flux ropes along PIL2, which decreased the local magnetic pressure,} led the lower flux rope along PIL2 to rise and expand, and to finally erupt as well (during flare2 and causing the associated CME2).  See Appendix~\ref{apd:11429_cme2} for some details of CME2. 

\section{Summary and Discussions}\label{sec:con}
 
In this paper, we analyze $188$ quasi-homologous CMEs with waiting times less than $18$ hours,  
and find that the waiting times show a Gaussian distribution peaking at about $7.5$ hours. Thus, the CMEs are believed to be physical related in the statistical sense. 
A classification based on the precise source locations has been performed:   
QH CMEs that sharing the source locations with their predecessors are defined as S-type, and the ones having different source locations from their predecessors are defined as D-type. Same source location means the involvement of the same part of a PIL
and different source locations mean different parts of one PIL or different PILs in an AR.  
In total, we classified $90$ S-type QH CMEs, and $52$ D-type ones.
Six cases, three of D-type and three of S-type, are discussed in Sec.~\ref{sec:source} to show the 
process of detailed identification, 
basically based on the corresponding localized flaring signatures such as ribbons and post-flare loops across the PILs.

The waiting time distributions of the two types of QH CMEs are significantly different: the distribution of the S-type CMEs peaks at around $7.5$ hours while the distribution of the D-type CMEs peaks at around $1.5$ hours,  suggesting that the major mechanisms of the two types of QH CMEs are probably different. 
In order to picture the differences in the possibly underlying mechanisms, one of S type and one of D type cases, are analysed in detail.
 
The S-type CME and its predecessor (i.e., CME$2$ and CME$1$) originated from the same location with a waiting time of $2.2$ hours in the quadrupolar AR 11158. 
Three parameters: the squashing factor $Q$ and the twist number $T_w$ that can locate the inner flux rope,   the decay index $n$ that measures the external confining force, are investigated at the erupting region at  Time$1$ (the time instance before the CME$1$) and Time$2$ (the time instance after CME$1$ but before CME$2$).    
The decay index above the erupting region shows no significant change, 
supporting that CME$1$ did not weaken the external confinement significantly. Note,  
the coronal magnetic field is extrapolated using the photospheric magnetograms as boundaries. 
It is possible that the change of the magnetic field in the corona cannot feed back to the photosphere within a short duration due to the high plasma $\beta$ (ratio of gas pressure to magnetic pressure) and the long response times of the photosphere relative to the corona, thus, the decay index remains unchanged. At both time instances, the height where decay index reaches the critical value for torus instability is much higher than the height of the flux rope axis, which suggests that torus instability may not be the direct causes for the two CMEs. 

The differences between the flux rope field lines that traced from the same starting coordinates in the pre- and post-CME1 corona indicates a topological change during flare1/CME1; while the reduction of the representation of the flux rope axial magnetic flux from Time1 to Time2 {\jkt evidence an eruption; presence of a BP and post-eruption loop at the position of the upper part of the flux rope at Time1 is more likely to support a partial expulsion process} : part of the flux rope erupted as CME1, while the other part may survive, erupting later as CME$2$, which fits into a free energy multi-stage release process. However, the magnetic free energy in the extrapolation volume almost remains unchanged, which may be due to three reasons:(i) the small extent of the CME-involved corona, small compared to the entire AR for which the energy budget was estimated, (ii) on-going free energy replenishment, (iii) the uncertainty of the free energy estimate itself. 


Besides the scenario of the S-type case in AR 11158, the eruptions from the same location can also be in a energy consuming and replenishment process 
as studied in \citet{rliu_2016}. Two CMEs with a waiting time of $13$ hours originated from the main PIL of a bipolar AR, AR 11817. The first one erupted and took the majority of the twist of the flux rope structure \citep[Fig.9 in][]{rliu_2016}. 
A very weakly twisted structure still existed after the eruption, 
and gained the twist through continuous shear motion on the photosphere \citep[Fig.10 in][]{rliu_2016}, and finally grew into a highly twisted seed flux rope for the next eruption. 
In this case, CME$1$ consumed most of the free energy at the erupting location, and the energy for CME$2$ was refilled after CME$1$. In the case of AR 11158, CME$1$ may only consumed part of the free energy,  
and the energy regain was ongoing before and after CME$1$ through the shear motion and flux emergence at the PIL \citep{Sun_2012a}. 
Although the amount of the consumed energy for CME$1$ may be different, they both are due to continuous energy input, fitting into the energy regain scenario. The BP of the flux rope in AR 11158 is probably the reason for preventing the flux rope 
from a full eruption whereas the 
rebuilding of magnetic free energy, e.g., flux emergence and shear motions, should be the main reason for the S-type eruptions. 
Detailed study of another CME-rich AR, AR 9236 that produced more than $10$ S-type CMEs with a mean waiting time around $7$ hours, also suggests that those S-type CMEs were caused by continuously emerging flux, supporting the free energy regain scenario \citep{Nitta_2001, Zhang_Wang_2002, Moon_2003b}.

The peak value around $7.5$ hours of the S-type QH CMEs waiting time distribution could be a characteristic time scale of the free energy replenishment process.

The D-type eruption and its predecessor originated from two different locations in AR 11429 with a waiting time of $1$ hour. No significant change is found in the decay index, like that in AR 11158. Again, the heights where decay index reaches the critical value for torus instability are much higher than the heights of the flux rope axes at both time instances, arguing against torus instability in triggering the two CMEs. However, the seed flux rope for CME$2$, i.e., the lower flux rope at PIL$2$ shows a stronger twist, clear rising and expansion after CME$1$, which are favourable for its eruption. The most possible reason for the change of the flux rope is that CME$1$ {\jkt influence the magnetic environment on PIL2 that 
make the upper two flux ropes 
disappear, lead to decrease of the local magnetic pressure 
and allow the lower one to erupt.  
In post-CME1 model corona, the upper two flux ropes totally disappeared from the domain. During flare$1$, a pair of ribbons ignited along PIL$2$ after the brightening of the ribbons along PIL$1$, with no development departing from the PIL, supporting a local reconnecting annihilation between the upper two flux ropes, rather than expulsion of them. The details about how the eruption of the flux rope along PIL$1$ resulted in the reconnection of the upper two flux ropes along PIL$2$ remains unclear, though the observation data has been analysed. The first CME can remove the common
overlying arcades, cause disturbance, compress the fields in neighbour system, 
even reconnect with neighbour fields. Somehow the equilibrium of the triple-decker flux rope is broken, and the upper two flux ropes reconnect. 
The key reason for the D-type eruption studied here is that the two flux rope systems are close enough that CME$1$ can impact on the pre-eruptive structure of CME$2$. It should be noted that the triple-decker flux rope presented here delivers a quiet uncommon configuration, of which equilibrium and evolution is worth to be studied in the future.} 


A well-studied D-type QH CME from AR 11402, with a waiting time of 48 minutes, also suggests that the CME was initiated 
by its predecessor 
\citep{Chengx_2013}. The first CME may have opened some overlying arcade, allowed the neighbouring fields to expand and lowered the downward magnetic tension above the neighbouring flux rope, leading to the second CME.
The scenario, that one eruption weakens the magnetic confinement of another flux tube system and promotes other eruptions, has been demonstrated in simulations \citep[e.g.,][]{Torok_2011, Lynch_2013}. The configuration in \citet{Torok_2011} 
contains a pseudo-streamer (PS), with two flux ropes located in the PS and one flux rope located next to the PS.   
The flux rope outside 
expands and erupts as the first CME, causing a breakout reconnection above one of the flux ropes in the PS, resulting the second CME; the current sheet formed below the second erupted flux rope causes reconnection at the overlying arcades of the other flux rope in the PS, leading to the third CME. The latter two CMEs can happen in a more generic configuration, without a flux rope outside the PS to eurpt at first to trigger them, 
although the underlying evolution is the same \citep[]{Lynch_2013}. The model of \citet{Torok_2011} or \citet{Lynch_2013} is applicable in a PS configuration. 
More generally, it is applicable in 
a configuration with 
a closed flux system containing a flux rope located nearby the erupting flux rope, e.g., a quadrupolar configuration, as the D-type CME and it's preceding one from AR 10030 shown in Fig.~\ref{fig:eru_10030}. 
The CME had a waiting time of $1$ hour, following a process similar as the second and third CMEs in \citet{Torok_2011}, or the two CMEs in \citet{Lynch_2013}, according to  
\citet{Gary_2004}: the core flux rope of the first CME was released from one flux tube system in a quadrupolar region by a breakout reconnection at the X point above the region; 
the neighbouring flux rope started to expand and finally erupted out due to the decrease of the overlying magnetic tension, which was caused by the reconnection at the current sheet formed below the first erupted flux rope. 
 
More generally, in an AR with multiple flux tube systems, one eruption causes destabilizations
that promote other eruptions could be described as a ``domino effect" scenario {\citep{Liuc_2009, Zuccarello_2009}. 
The peak value of the waiting time distribution of the D-type QH CMEs, around $1.5$ hour, could be the characteristic time scale of the growth of distablization 
caused by their predecessors. 
This kind of consecutive CMEs 
with extremely short waiting time are sometimes called as ``twin-CMEs'' or ``sympathetic-CMEs", although they are not necessarily produced from the same AR  \citep[e.g.,][]{Schrijver_2011, Balasubramaniam_2011, Yang_2012, Shenc_2013, Ding_2013, Ding_2014}. 
The source locations of a D-type QH CME and its predecessor 
are expected to be located close to each other, or have some magnetic connection that one eruption can induce the other one.   

Note, there is another slightly lower peak around $9.5$ hours in the  waiting time distribution of D-types, may be due to the method of classification, or even different mechanism from the one for those with waiting time around $1.5$ hours. 

In conclusion, through the two cases studied in depth, we propose possible mechanisms for most of the two types of QH CMEs, i.e., the ones located around the peak of the waiting time distribution:  
S-type QH CME can occur in a recurring energy release process by free energy regain, while D-type QH CME can happen 
when disturbed by its preceding one. 
The different peak values of the waiting time distributions: $7.5$ hours for S-type and $1.5$ hour for D-type QH CMEs might be the characteristic time scales of the two different scenarios. 
The classification is only based on the source PILs. S-type QH CME may also happen when disturbed by its predecessor, following a process as similar to the D-type. For example, in a configuration with {\jkt more than one flux ropes vertically located above the same PIL, like the ones in AR 11429, in which change (reconnection, expulsion, etc.) of the upper} flux ropes caused the eruption of the lower one. 
More cases with high spatial and temporal resolution data (e.g., data from $SDO$) are worth to be studied to discover more scenarios.

\clearpage
 
\begin{figure*}
\begin{center}
\includegraphics[width=0.45\hsize]{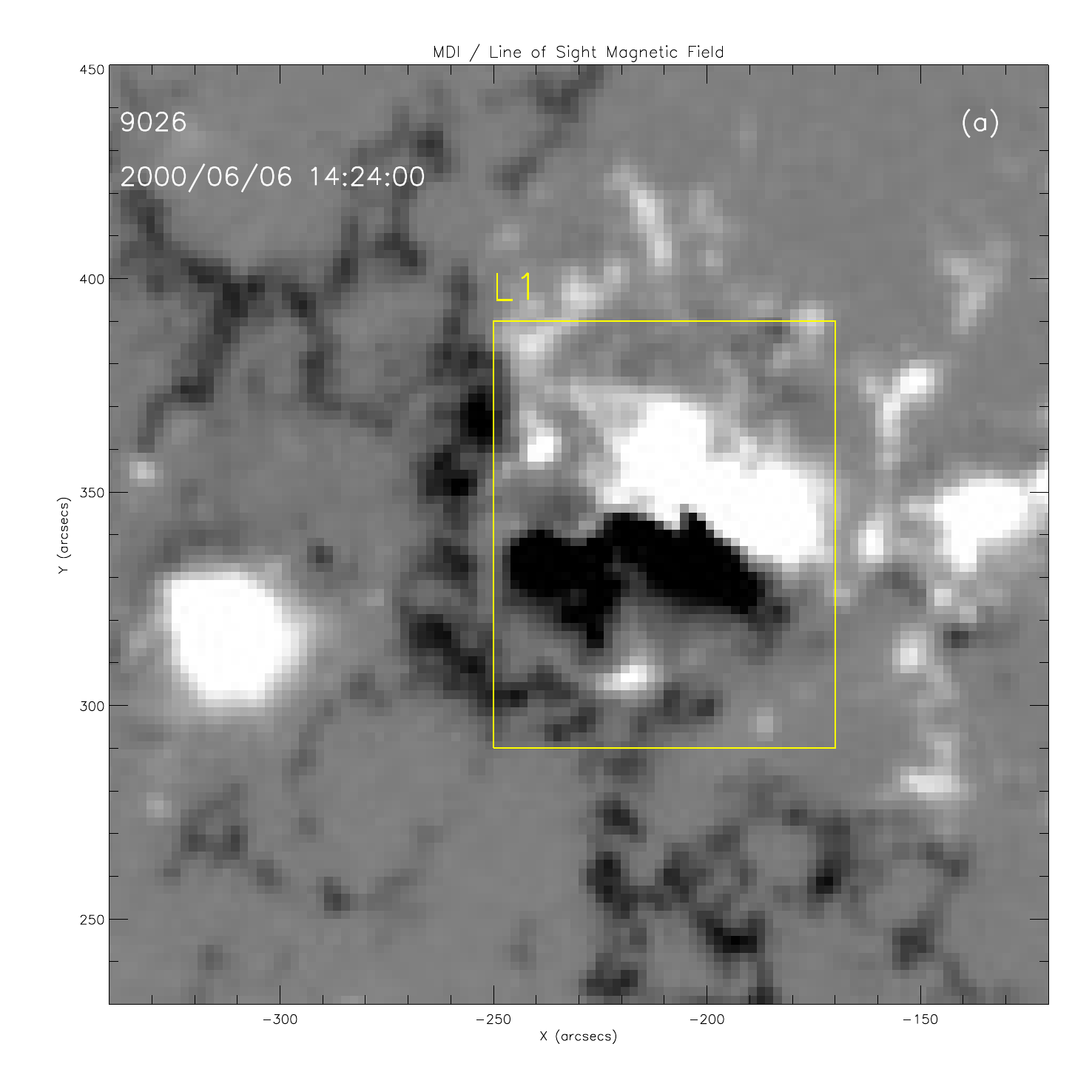}
\includegraphics[width=0.95\hsize]{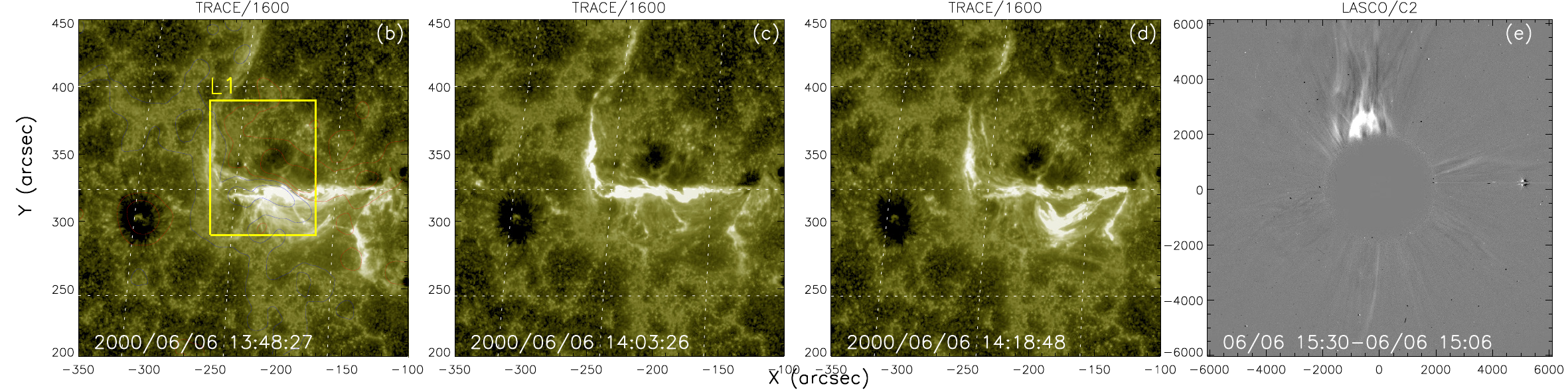}
\includegraphics[width=0.95\hsize]{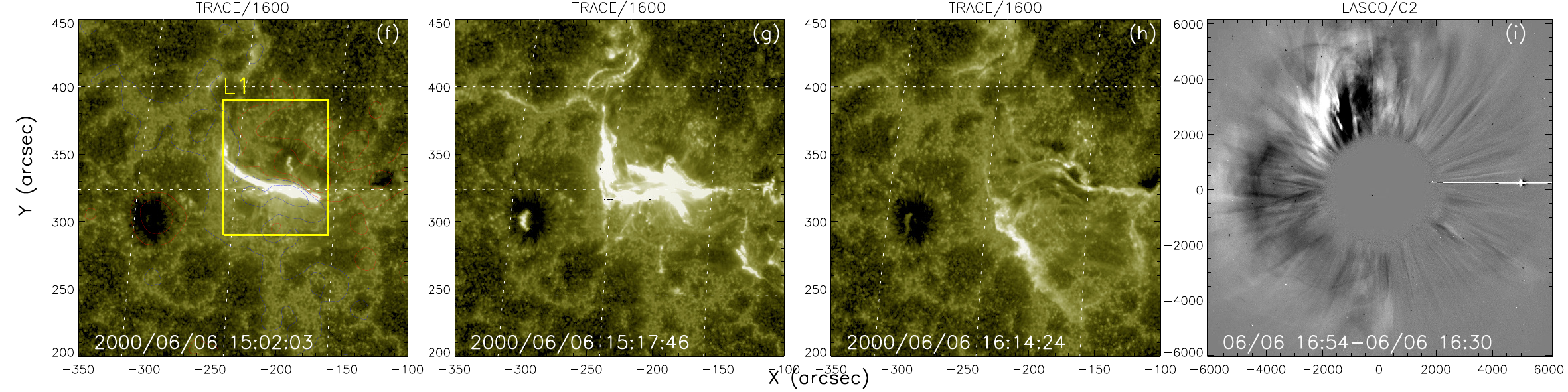}
\caption{An S-type CME and its predecessor that both originated from AR 9026. (a) {\it SOHO}/MDI photospheric LOS magnetic field. Black/white color represents negative/positive magnetic polarity. The yellow box L$1$ outlines the source location identified for the two CMEs.
Panels (b)--(d) and (f)--(h) show the chromospheric flaring features associated to the preceding and following CME, 
respectively. Red and blue contours in (b) and (f) are drawn at $\pm[150,850]\,G$, respectively. Panels (e) and (i) show the white-light signatures of the two QH CMEs. See also the corresponding online animation. 
} 
\label{fig:eru_9026}
\end{center}
\end{figure*}

\begin{figure*}
\begin{center}
\includegraphics[width=0.45\hsize]{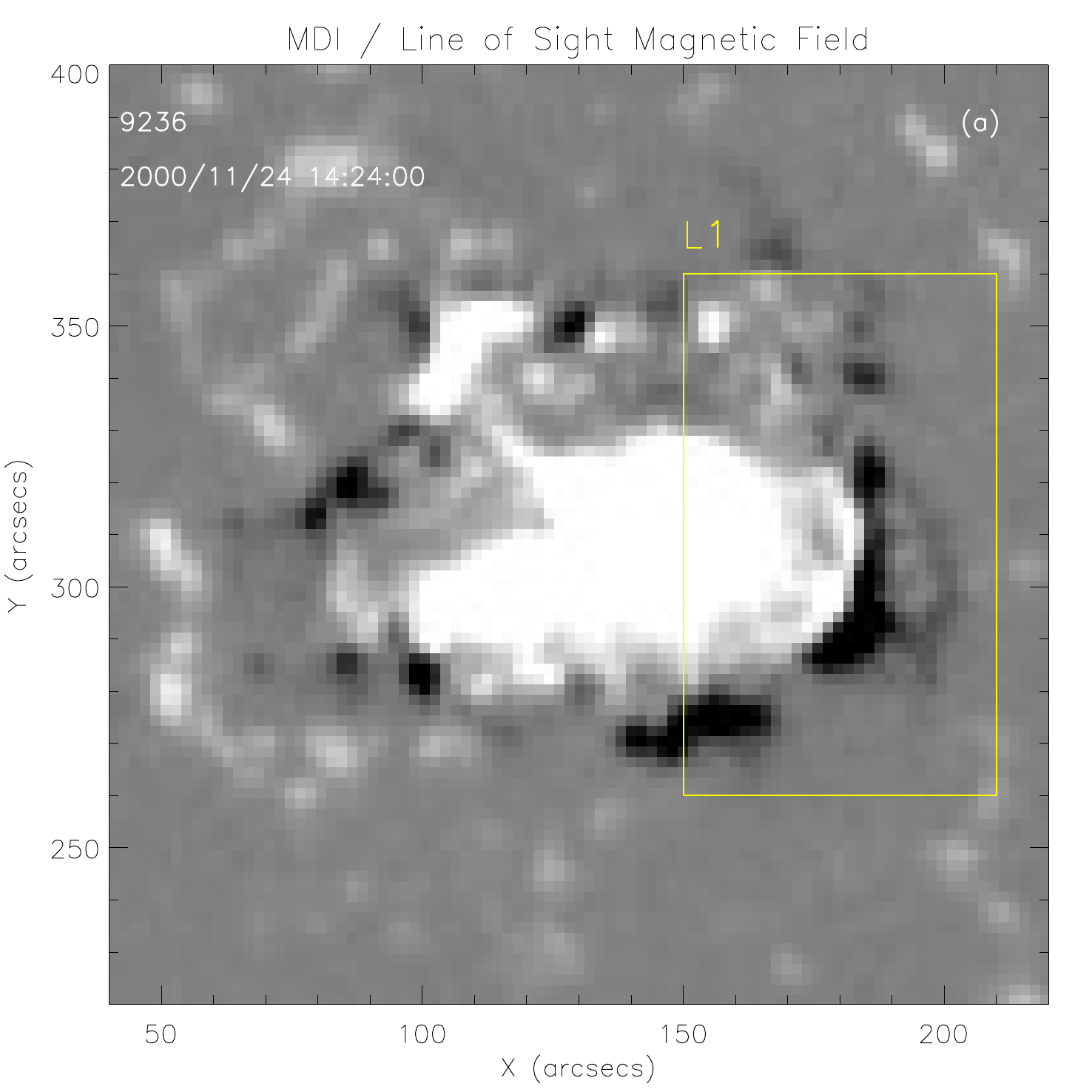}
\includegraphics[width=0.95\hsize]{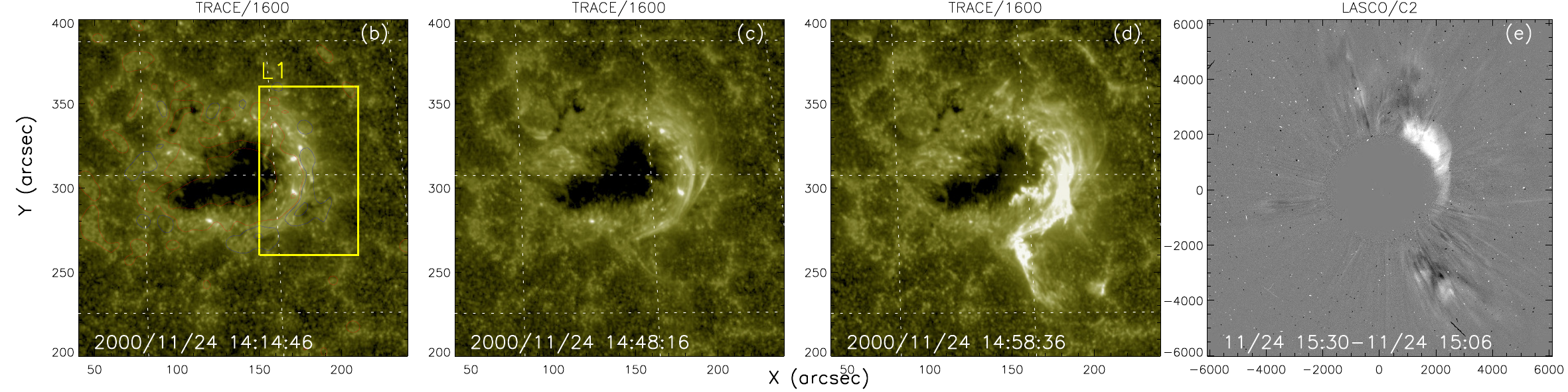}
\includegraphics[width=0.95\hsize]{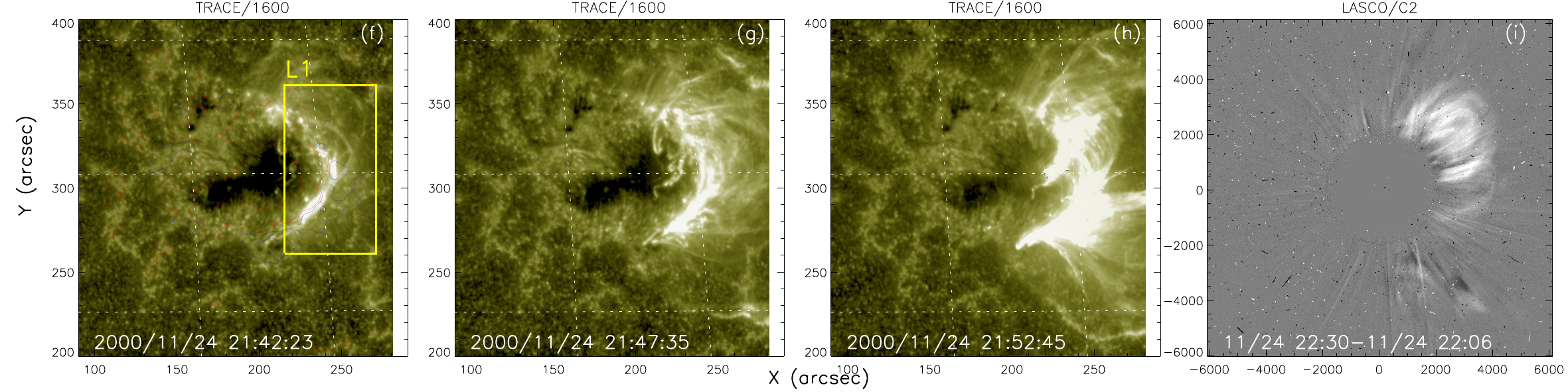}
\caption{An S-type CME and its predecessor from AR 9236. 
Same layout as Fig.~\ref{fig:eru_9026}. See also the corresponding online animation.}
\label{fig:eru_9236}
\end{center}
\end{figure*}

\begin{figure*}
\begin{center}
\includegraphics[width=0.85\hsize]{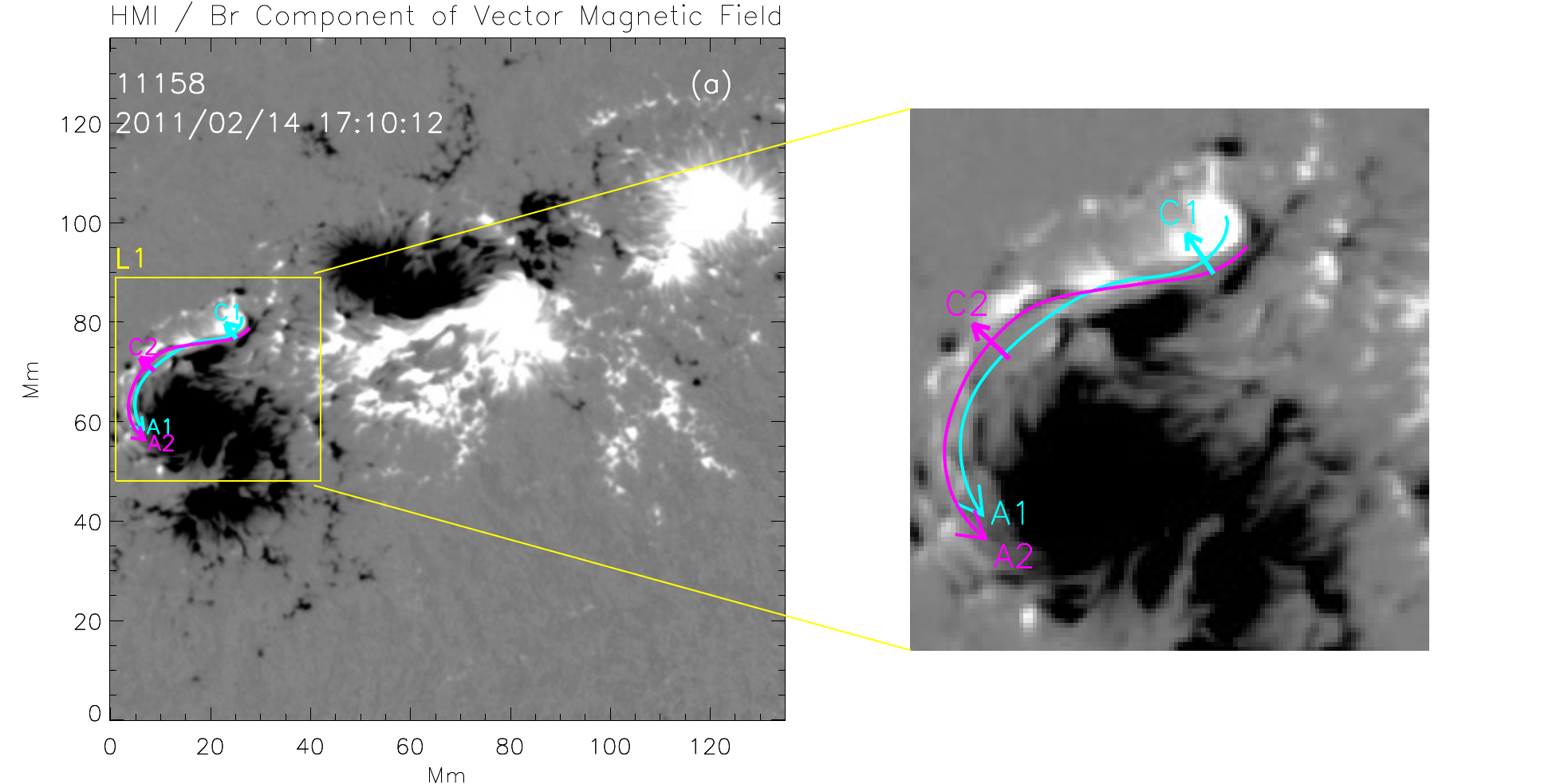}
\includegraphics[width=0.95\hsize]{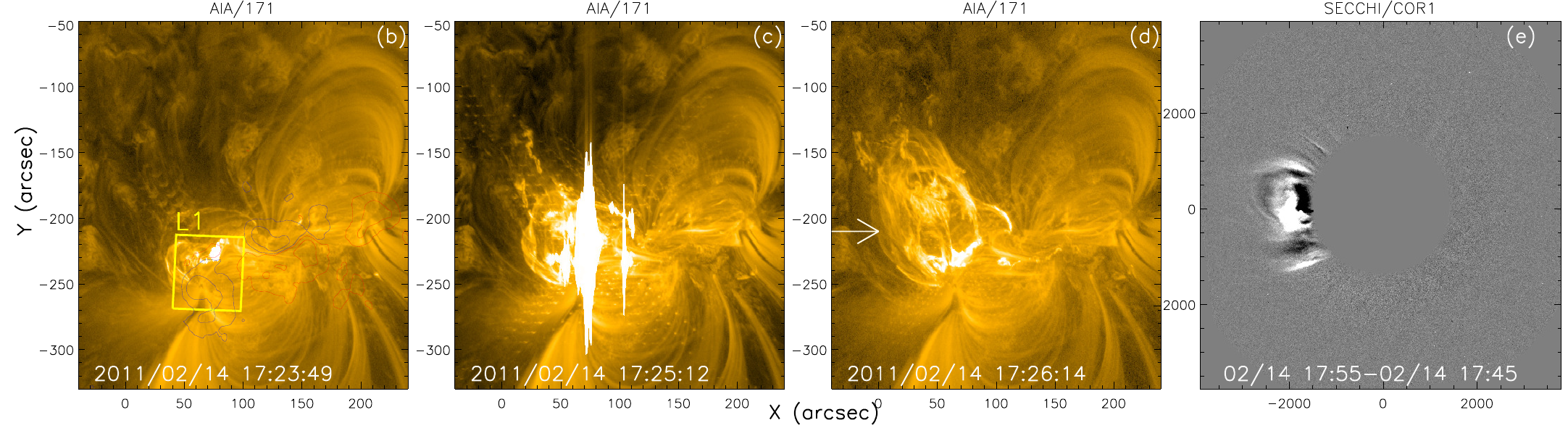}
\includegraphics[width=0.95\hsize]{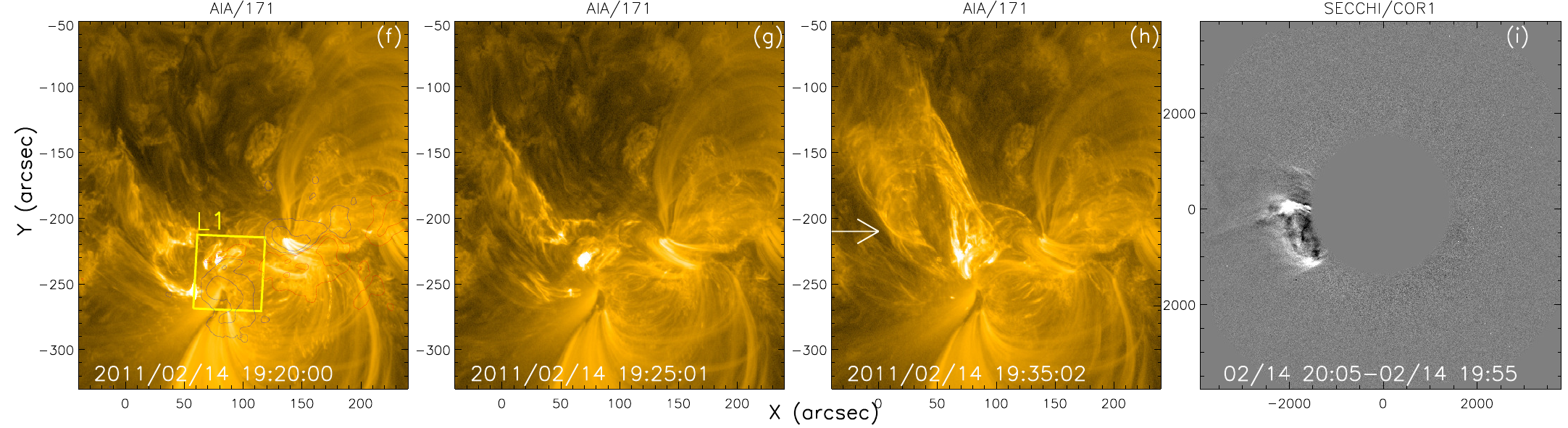}
\caption{
An S-type QH CME and its prodecessor that originated from AR NOAA~11158. Same layout as Fig.~\ref{fig:eru_9026}. 
The source location L$1$ in (a) is enlarged and shown in the right top panel. 
Panels (b)--(d) and (f)--(h) show {\it SDO}/AIA 171~\AA\ observations of the associated flares. 
The white arrows in (d) and (h) indicate the erupting mass of the two CMEs. 
Panels (e) and (i) show running-difference {\it STEREO}/COR1 images. The colored lines, labeled A$1$ and A$2$ in panel (a), 
outline the orientation of the axes of the magnetic flux ropes, that erupted to produce the associated CMEs.
C$1$ and C$2$ mark the footprints of two vertical planes
used to visualize the topological properties 
of the involved magnetic structures. Cyan and pink color represent the configurations at Time$1$ and Time$2$, respectively. See also the corresponding online animation.}
\label{fig:eru_11158}
\end{center}
\end{figure*}

\begin{figure*}
\begin{center}
\includegraphics[width=0.45\hsize]{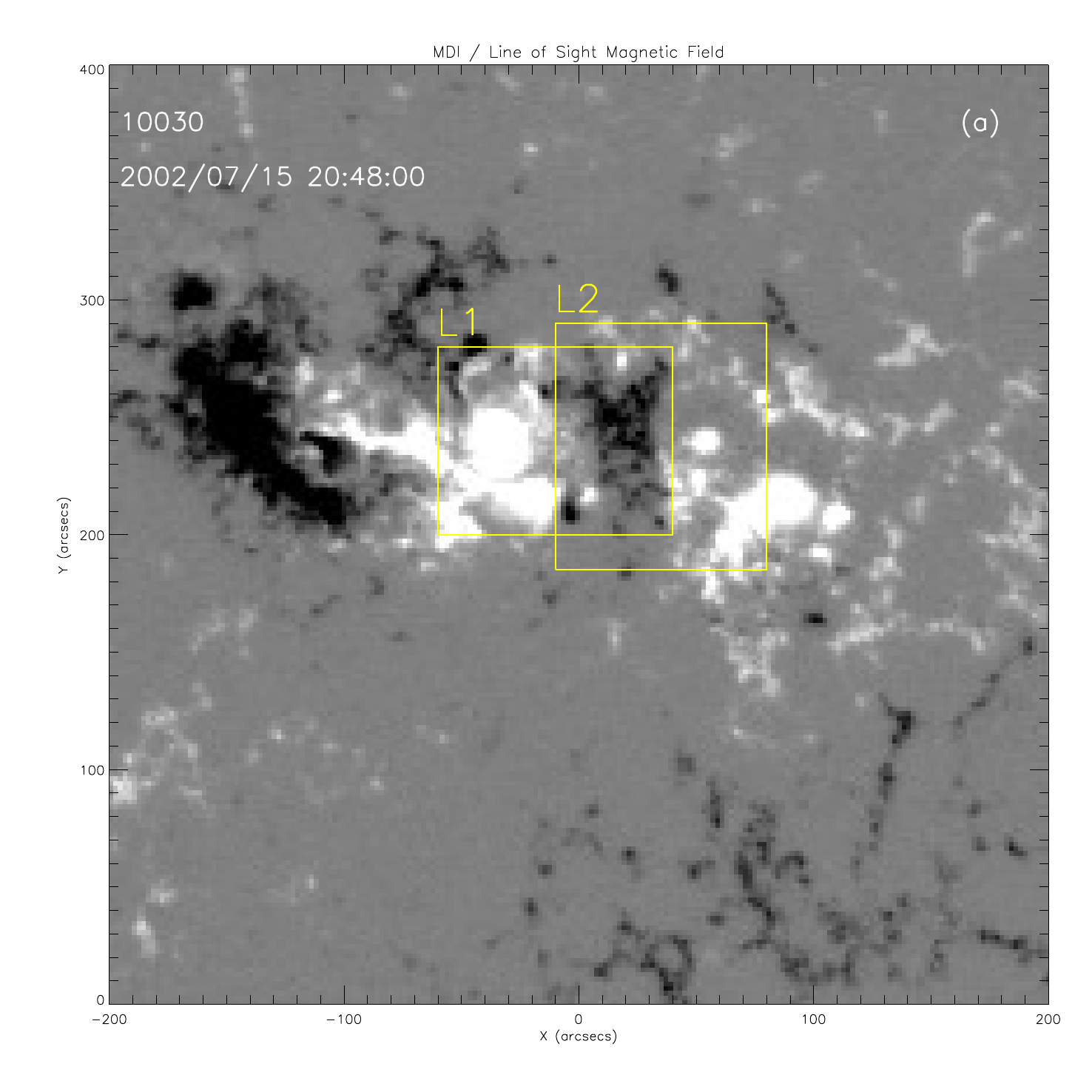}
\includegraphics[width=0.95\hsize]{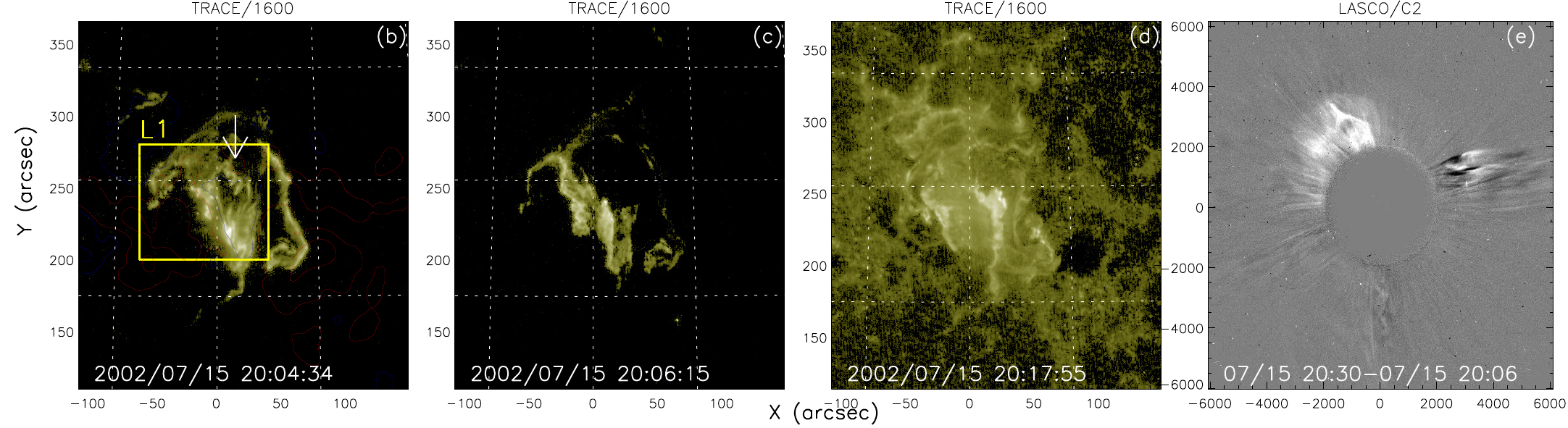}
\includegraphics[width=0.95\hsize]{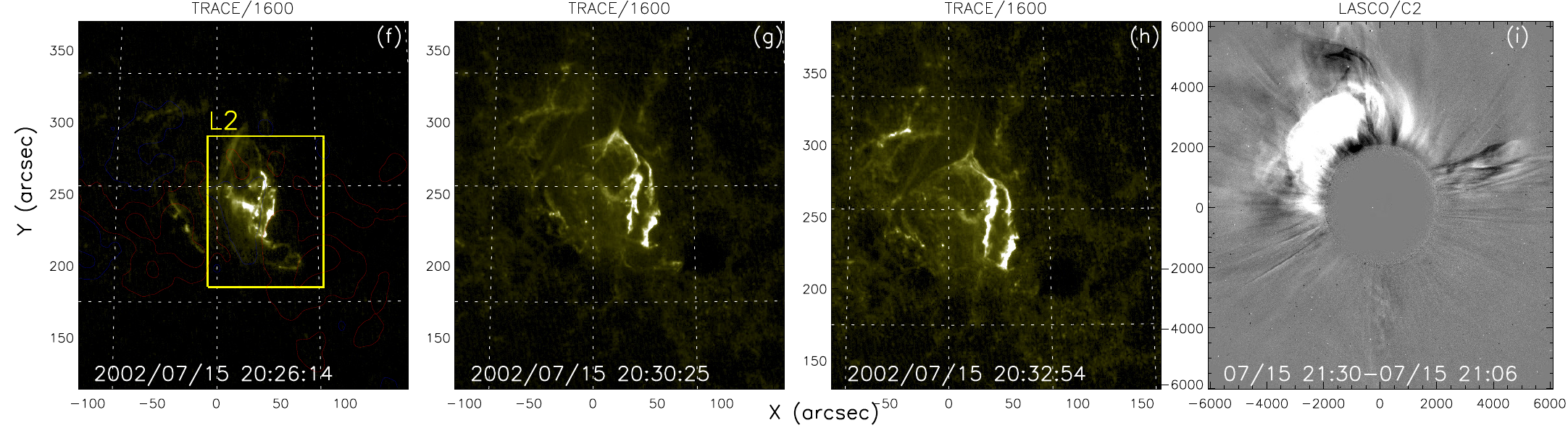}
\caption{A D-type CME and its predecessor that originated from AR NOAA~10030. Same layout as in Fig.~\ref{fig:eru_9026}. The yellow boxes L$1$ and L2 in (a) outline the source location of CME$1$ and CME$2$, respectively. The white arrow in (b) marks an erupting helical structure. See also the corresponding online animation. 
} \label{fig:eru_10030}
\end{center}
\end{figure*}

\begin{figure*}
\begin{center}
\includegraphics[width=0.45\hsize]{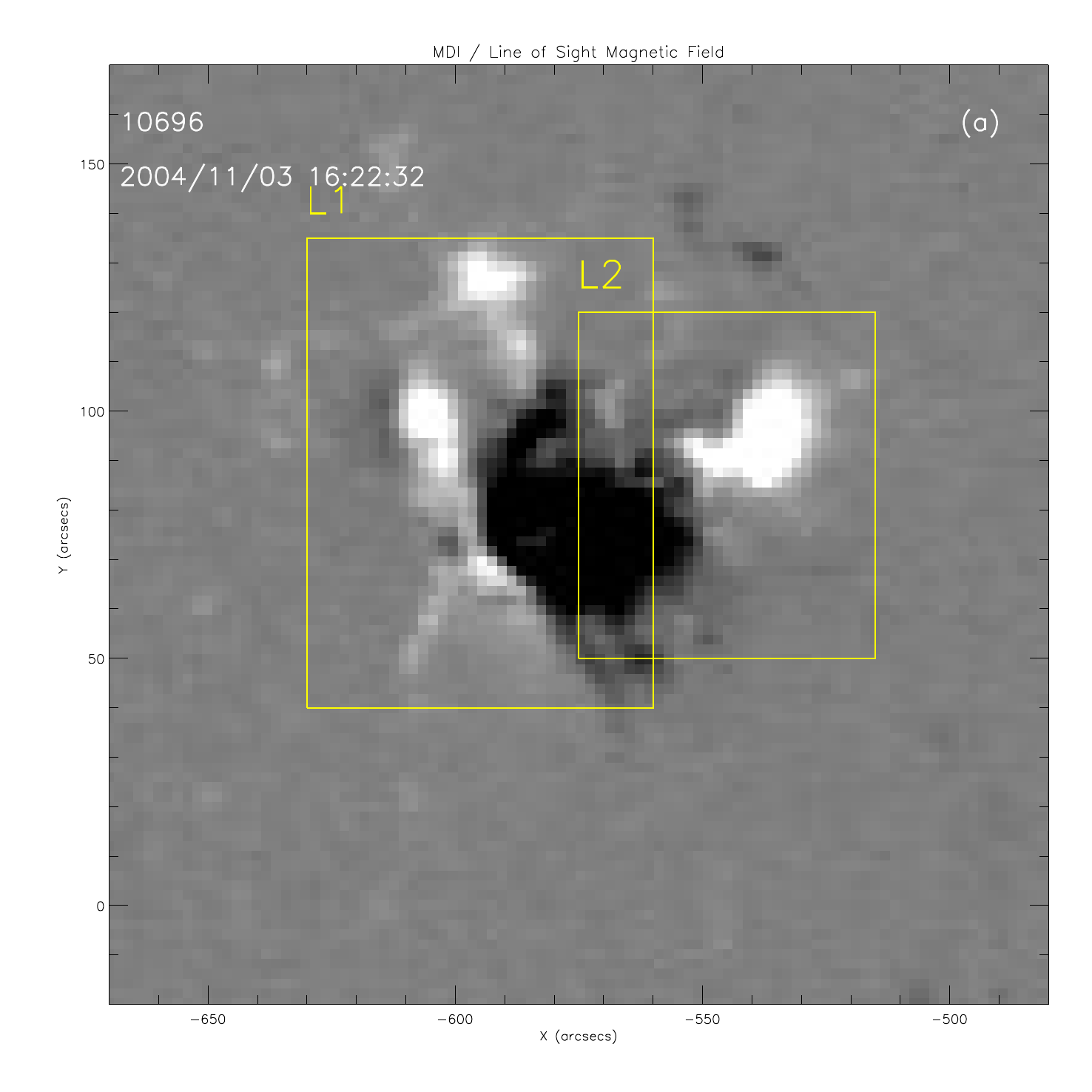}
\includegraphics[width=0.95\hsize]{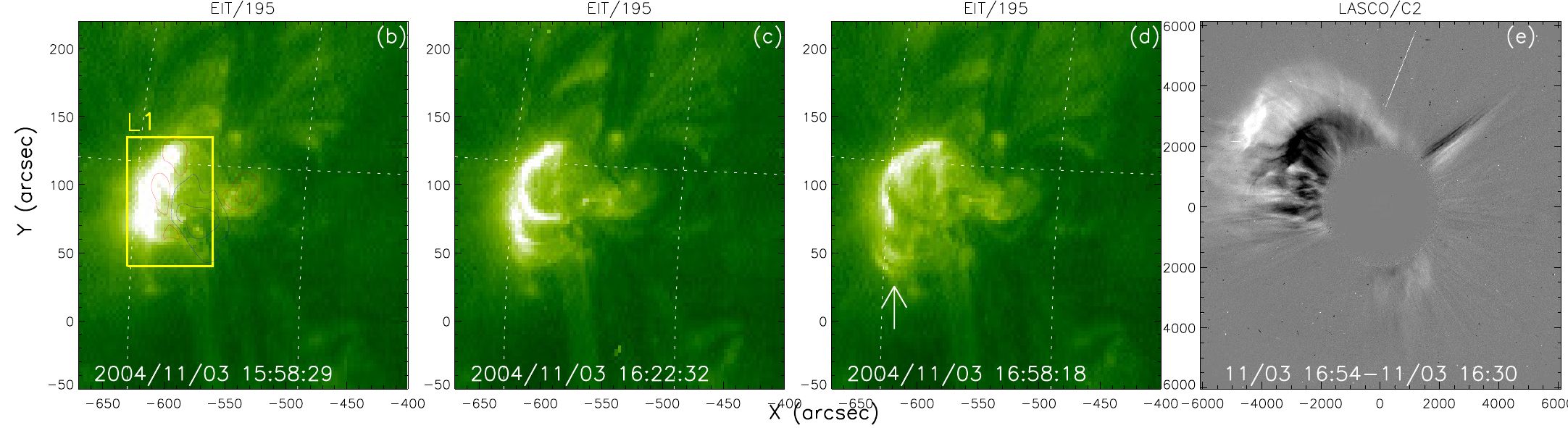}
\includegraphics[width=0.95\hsize]{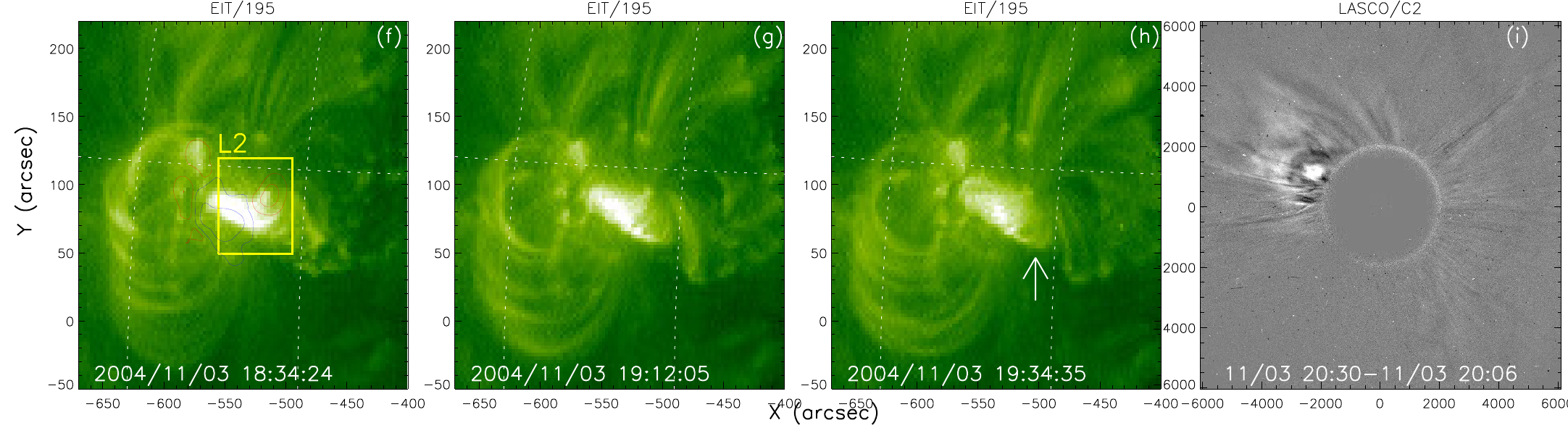}
\caption{A D-type CME and its predecessor from AR NOAA~10696. Same layout as in Fig.~\ref{fig:eru_10030}. Panels (b)--(d) and (f)--(h) show the flaring 
features associated with the first and second CME, 
respectively, as observed by {\it SOHO}/EIT at 195\AA. The white arrows in (d) and (h) indicate the post-flare loops associated with flare1 and flare2. See also the corresponding online animation.} \label{fig:eru_10696}. 
\end{center}
\end{figure*}

\begin{figure*}
\begin{center}
\includegraphics[width=0.85\hsize]{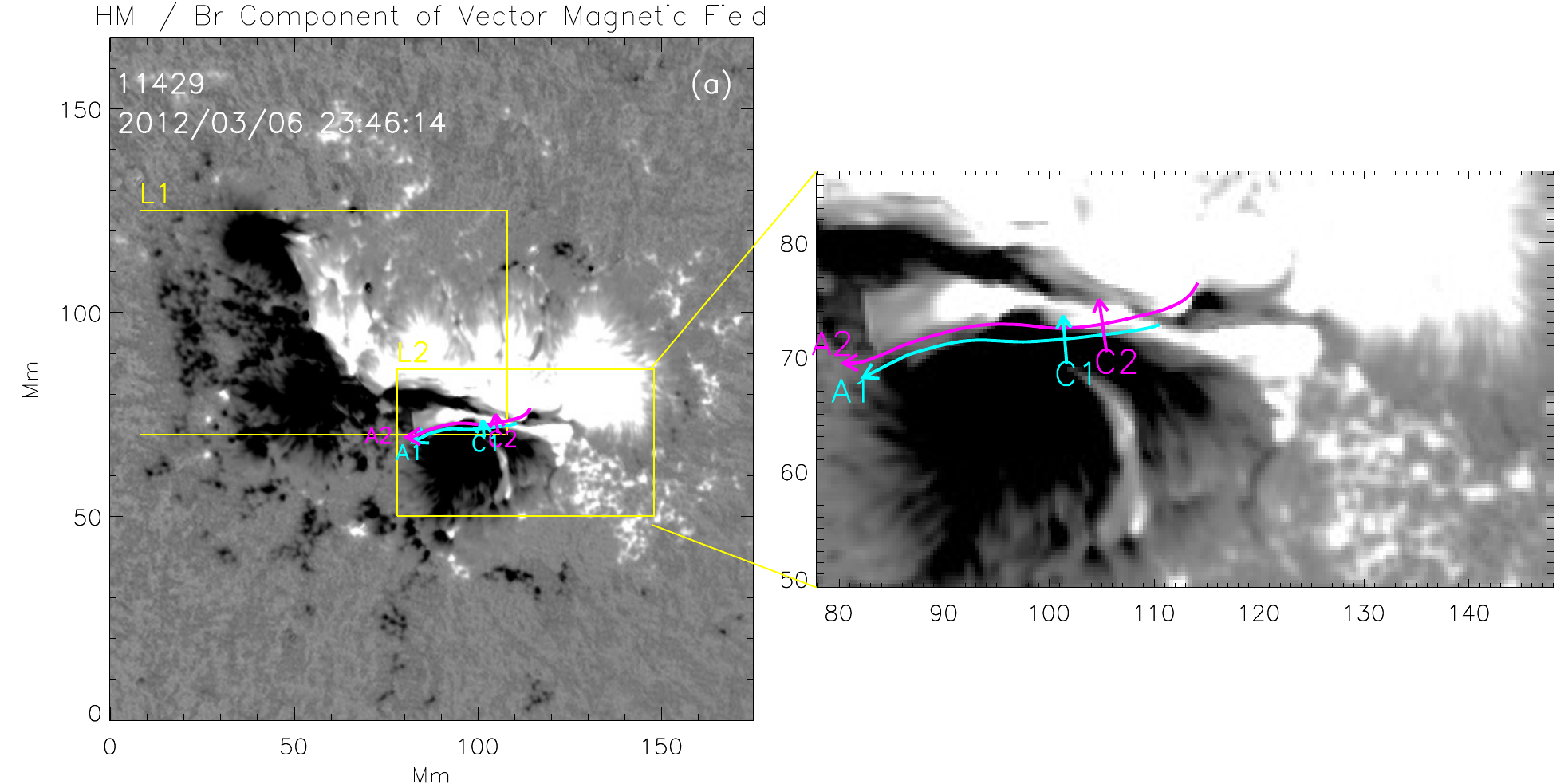}
\includegraphics[width=0.95\hsize]{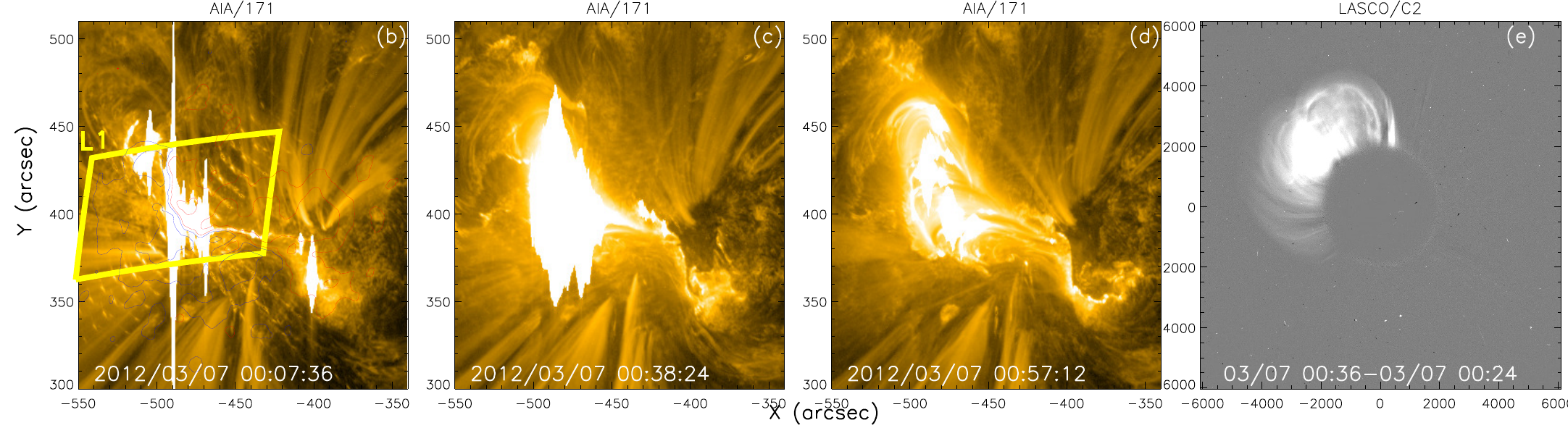}
\includegraphics[width=0.95\hsize]{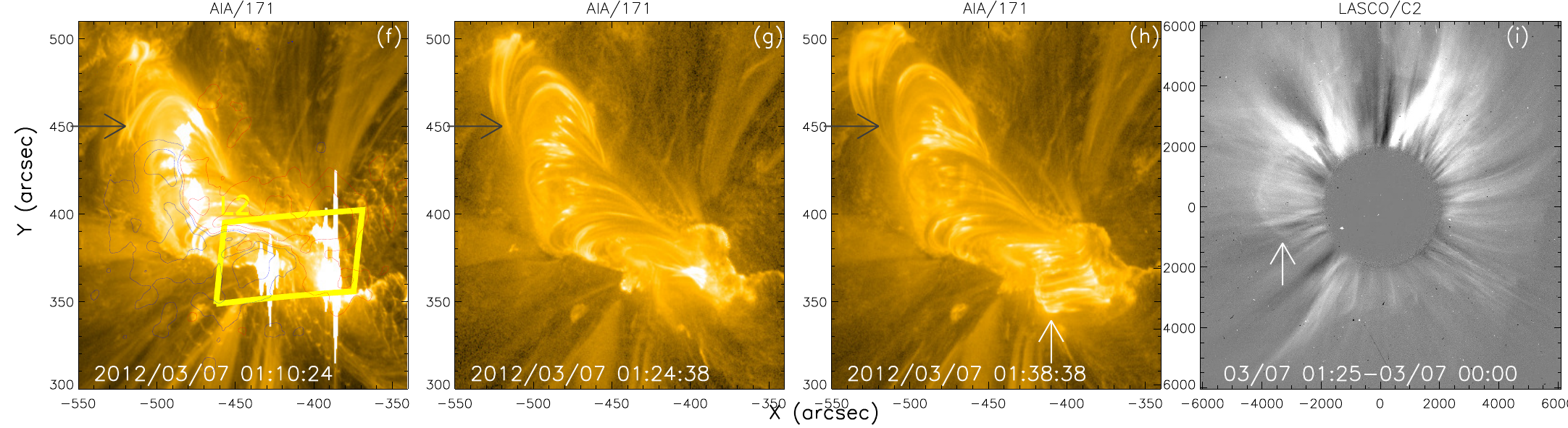}
\caption{A D-type CME and its predecessor from AR NOAA~11429. Same layout as in Fig.~\ref{fig:eru_10030}. The colored lines and arrows in panel (a) have the same meaning as the ones in Fig.~\ref{fig:eru_11158}. Panels (b)--(d) and (f)--(h) show the corresponding flaring features observed by {\it SDO}/AIA at 171\AA. The white arrow in panel (h) marks the post-flare loops of flare$2$, while the one in panel (i) marks the faint front of CME$2$. The black arrows in panels (f)--(h) mark the afterglow of flare$1$. See also the corresponding online animation.} \label{fig:eru_11429}
\end{center}
\end{figure*}

\begin{figure*}
\begin{center}
\includegraphics[width=0.75\hsize]{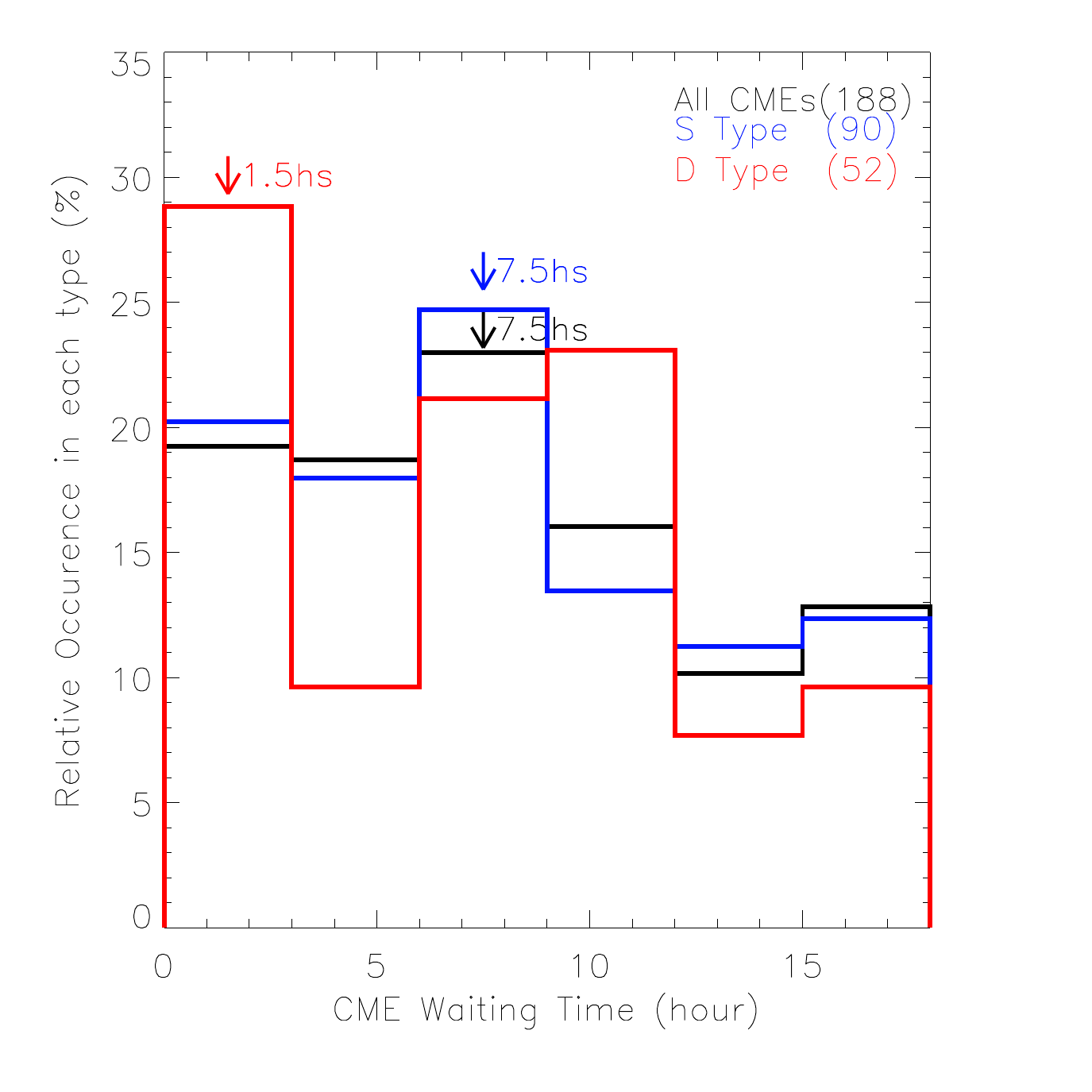}
\caption{Waiting time distributions for the 188 QH CMEs under study, exhibiting waiting times of $<$18~hours (black line). 
{The blue and red line represents the respective waiting time distributions for the S- and D-type events. Digits in brackets denote the number of QH CMEs in the corresponding sample.} Vertical arrows indicate the peak in the respective distribution.} 
\label{fig:wt}
\end{center}
\end{figure*}

\begin{figure*}
\begin{center}
\includegraphics[width=0.68\hsize]{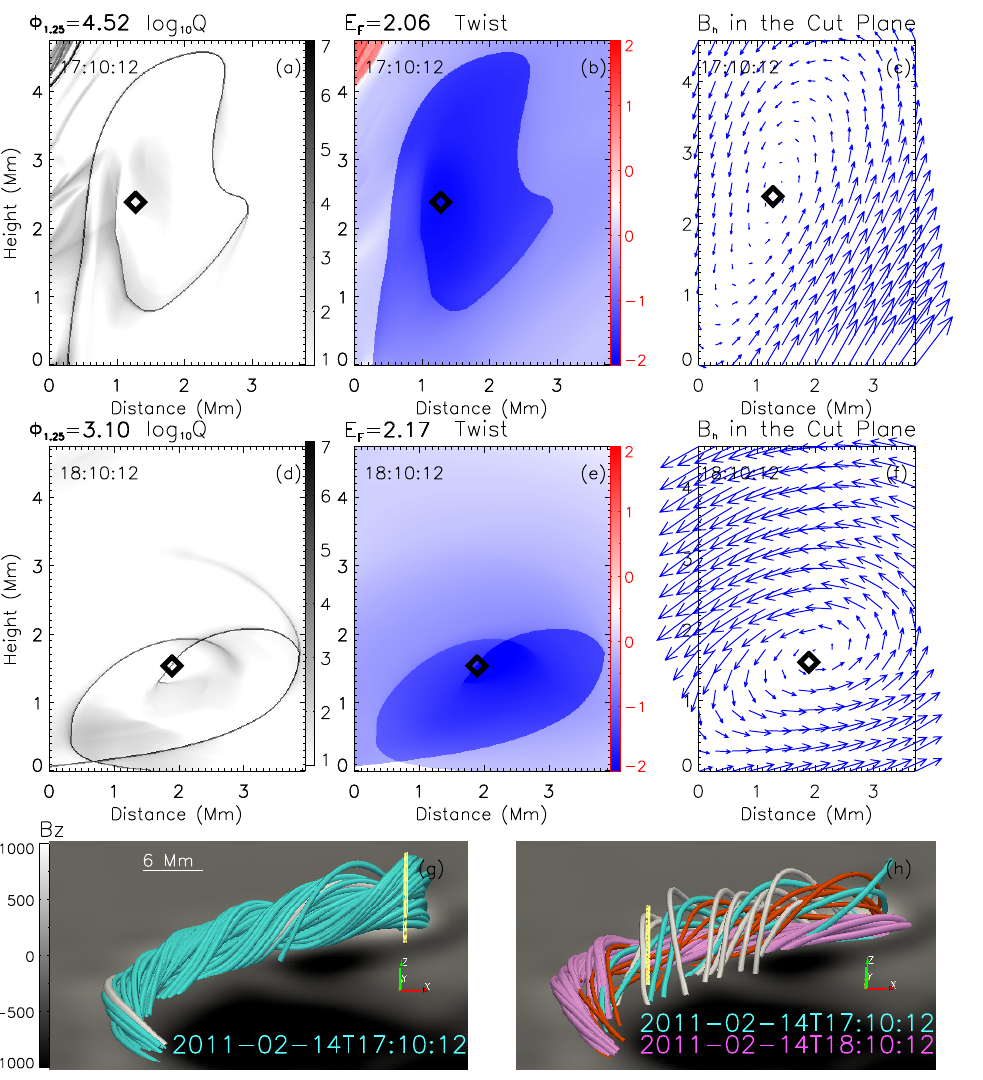}
\caption{Pre- (at Time1) and post-CME1 (at Time2) conditions in AR NOAA~11158. Panels (a), (b) and (c) show $Q$, $T_w$ and $\vec{B}_\parallel$ in a vertical plane perpendicular to the pre-eruptive flux rope axis. The footprint of the plane is indicated by the colored line C1 in Fig.~\ref{fig:eru_11158}(a). Panels (d) - (f) show the distribution of the same quantities at Time$2$, in a plane perpendicular to the flux rope axis (C2 in Fig.~\ref{fig:eru_11158}(a)). The yellow lines in panels (g) and (h) mark the positions and extents of the vertical planes. The blue arrows in (c) and (f) indicate the vector fields with the normal components going into the plane. 
The black diamonds in (a) - (f) mark the position where $T_w$ has its maximum. Panels (g) and (h) show the twisted field lines, traced based on the geometrical information in the $Q$ and $T_w$ maps. Cyan and pink lines mark the flux rope field lines at Time1 and Time2, respectively. The white line in (g) indicates a representative field line in the Bald Patch, while the white field lines in (h) show the arcade traced in the post-CME1 corona, but from exactly the same coordinates of the upper part of the flux rope at Time$1$. The cyan field lines in (h) roughly outline the flux rope at Time$1$ for comparison. The red lines in (h) are also some pre-CME1 flux rope field lines, but traced exactly from the coordinates of the high $T_w$ region at Time2 (panel(e)). $\Phi_{1.25}$, given at the headers of panel (a) and (d), are vertical magnetic flux (in unit of $10^{19} $\,Mx) from the strong $T_w$ region ($|T_w|\gtrsim1.25$) at each time, respective. $E_F$, given at {\jkt the headers of} panel (b) and (e), are free magnetic energy (in unit of $10^{32} $\,erg) at the two times. {\jkt The grey-scale bar at the left of panel (g) shows the scale of the photospheric magnetic fields plotted in panel (g) and (h), in unit of Gausses.} 
} \label{fig:rope_11158}
\end{center}
\end{figure*}

\begin{figure*}
\begin{center}
\includegraphics[width=0.95\hsize]{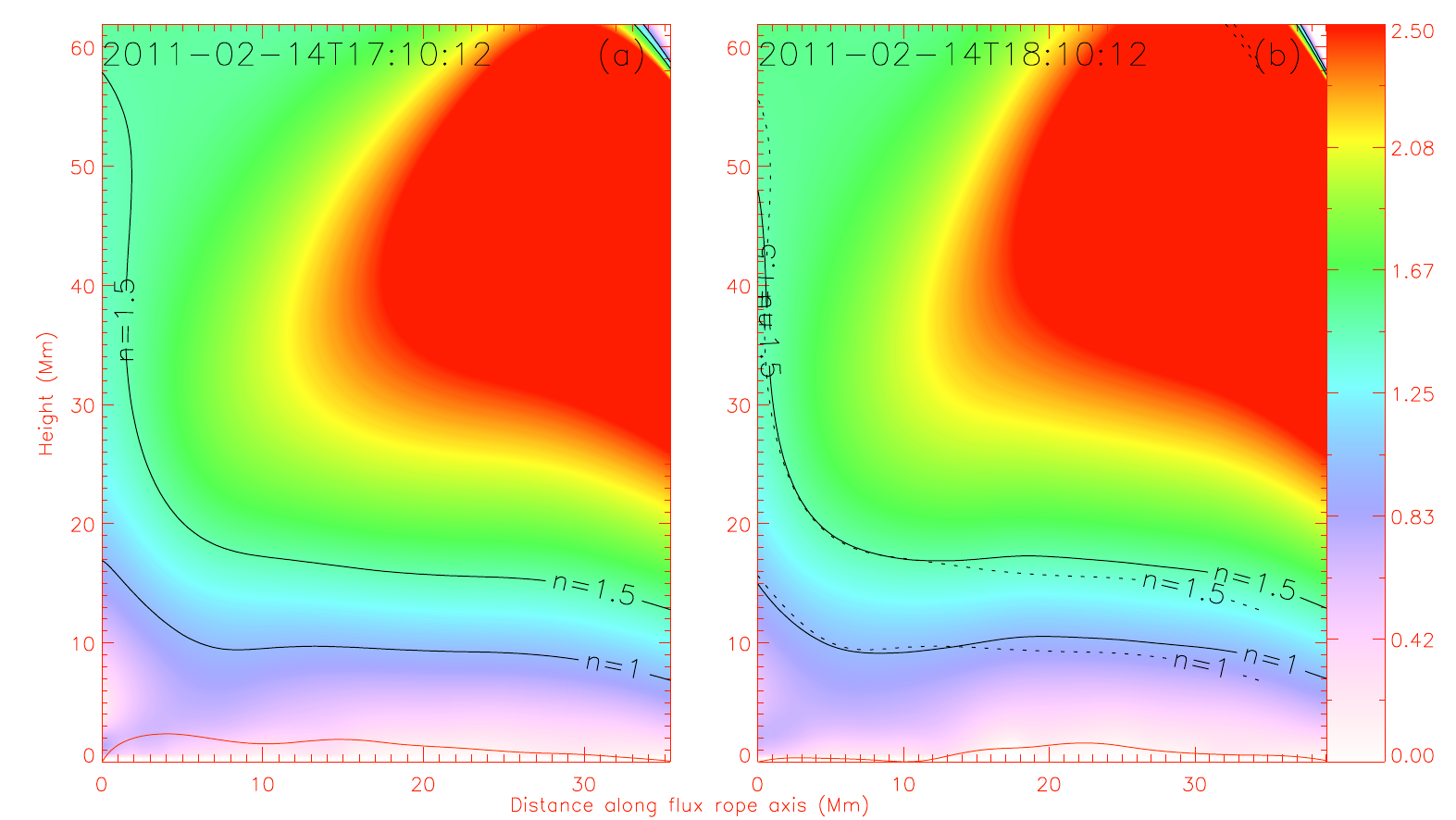}
\end{center}
\caption{ Vertical distributions of the decay index, $n$, above the axis of the (a) pre-CME1 and (b) post-CME1 flux ropes in AR NOAA~11158.
The black lines in (a) and (b) mark the height where $n=1$ and $n=1.5$ at the different time instances. The dotted lines in (b) mark the corresponding heights at Time$1$ for comparison. The red lines indicate the respective height of the flux rope axis. 
} \label{fig:decay_11158}
\end{figure*}

\begin{figure*}
\begin{center}
\includegraphics[width=0.8\hsize]{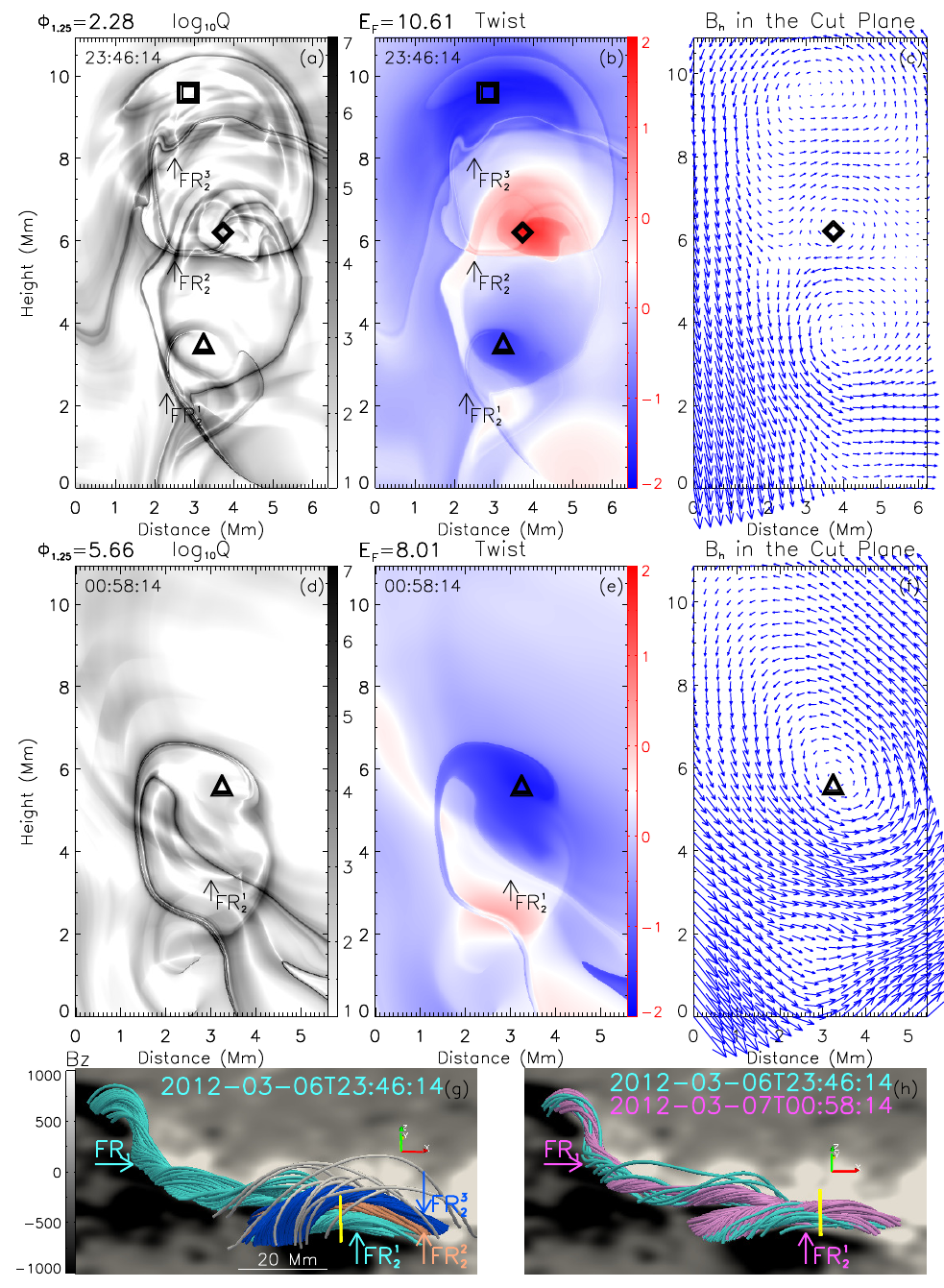}
\caption{Magnetic features in the vertical cuts (indicated by the colored cuts C1 and C2 in Fig.~\ref{fig:eru_11429} (a)) above the PIL$2$ in AR 11429 at Time$1$ and Time$2$. Same layout as Fig.~\ref{fig:rope_11158}. Arrows in panels (g), (h) mark the flux rope along PIL$1$ as FR$1$, {\jkt the lower (middle, upper) flux rope along PIL$2$ as FR$_2^1$ (FR$_2^2$, FR$_2^3$)}, same meaning in (a), (b), (d), (e). Yellow vertical lines in (g) and (h) mark the position of the vertical cuts. The white lines in (g) are some nearly-potential arcades above the flux ropes. The cyan lines in (h) roughly outline the flux ropes at Time$1$. $\Phi_{1.25}$, given at the headers of panel (a) and (d), are vertical magnetic flux (in unit of $10^{19} $\,Mx) from the strong 
$T_w$ region ($|T_w|\gtrsim 1.25$) of the lowermost rope at each time, respective. {\jkt $E_F$, given at the headers} of panel (b) and (e), are free magnetic energy (in unit of $10^{32} $\,erg) at the two times. } \label{fig:rope_11429}
\end{center}
\end{figure*}

\begin{figure*}
\begin{center}
\includegraphics[width=0.95\hsize]{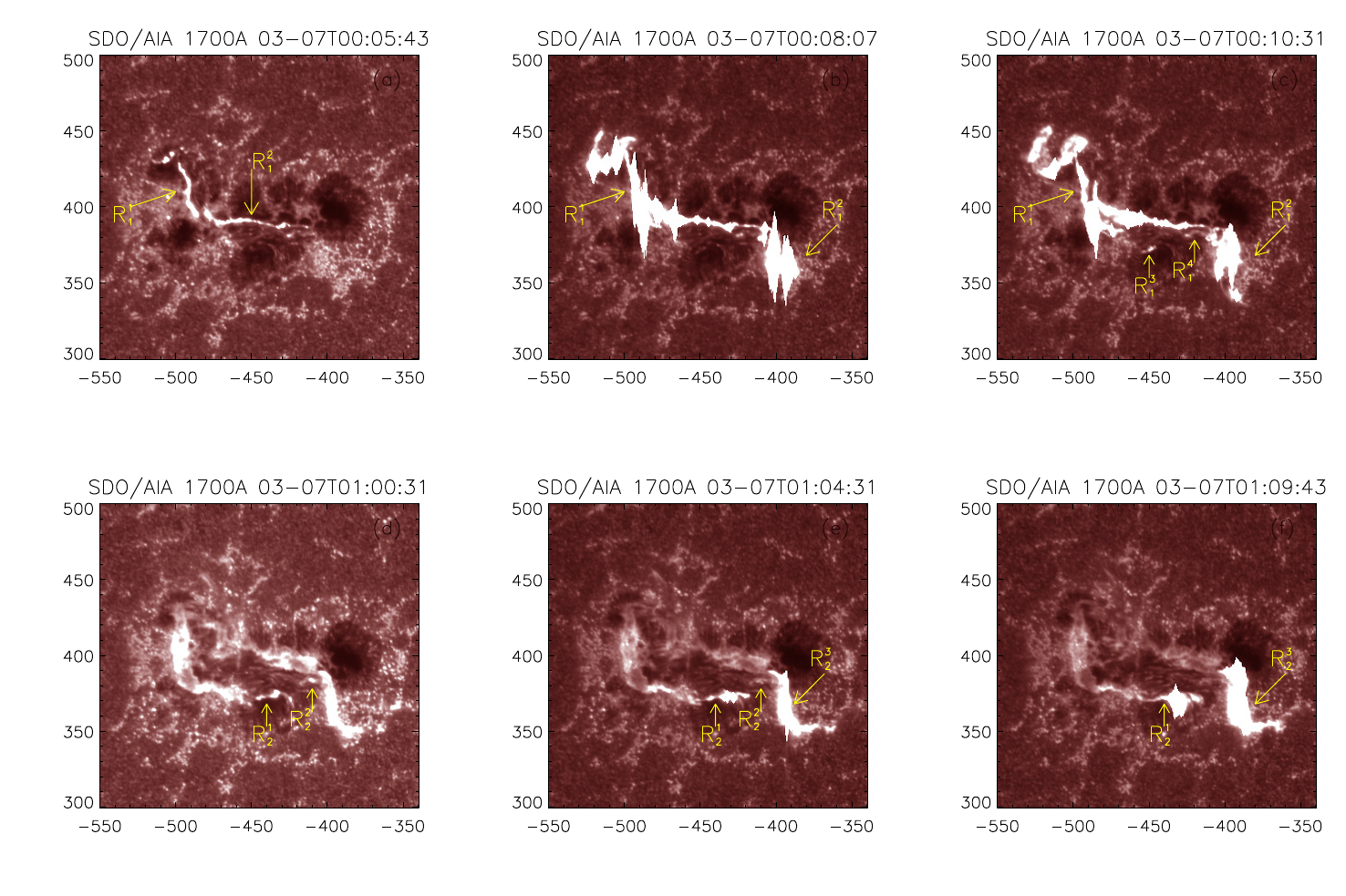}
\caption{Evolution of flare ribbons during the two QH eruptions in AR~11429. Panels (a) - (c) for flare$1$ and panels (d) - (f) for flare$2$. Yellow arrows mark different ribbons during the flares. R$_i^j$ denotes the $j_{th}$ ribbon for the $i_{th}$ flare. 
} \label{fig:rib_11429}
\end{center}
\end{figure*}

\begin{figure*} 
\begin{center}
\includegraphics[width=0.95\hsize]{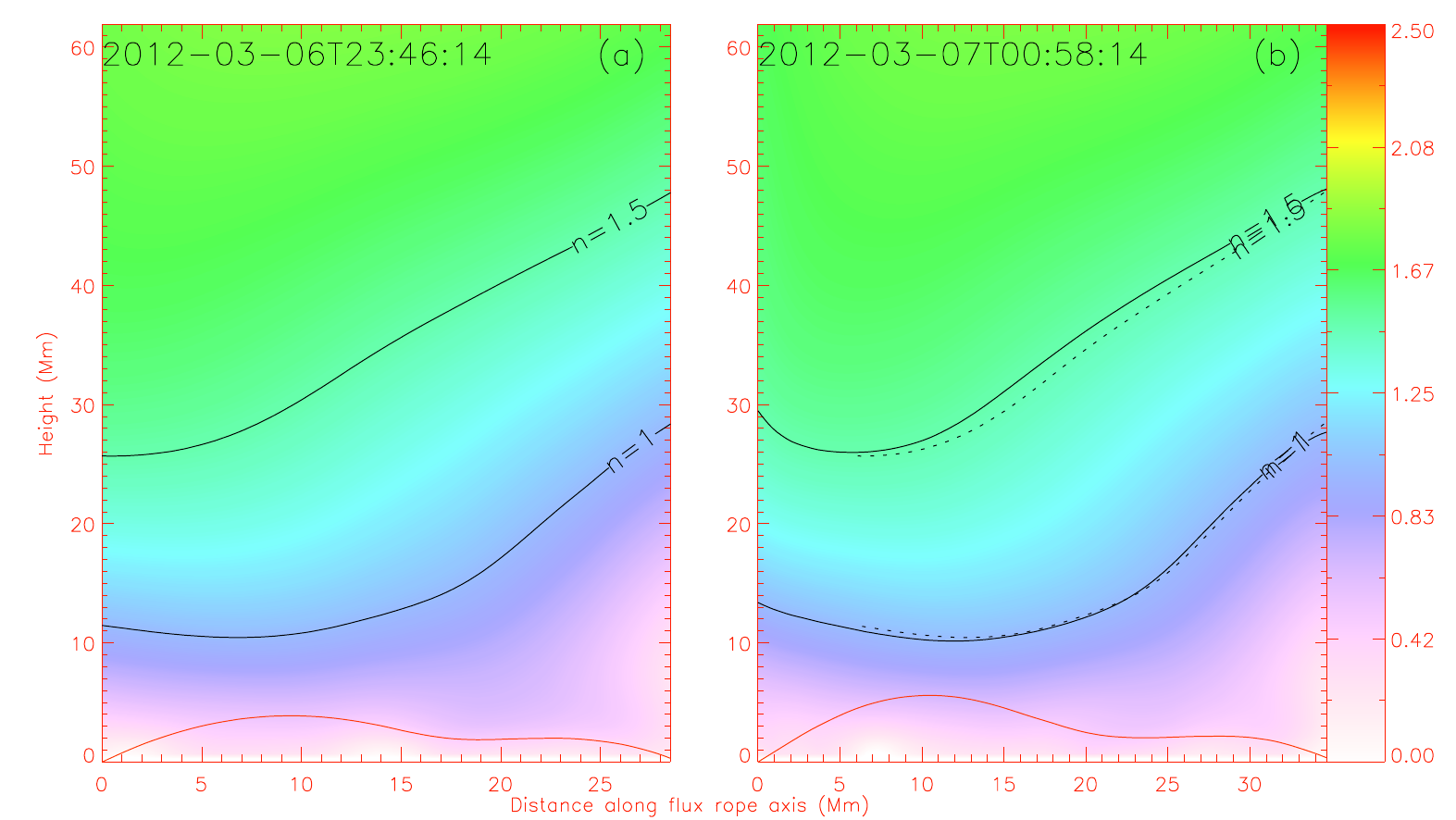}
\caption{Decay index distributition above the axis of the lower flux rope along PIL$2$ for AR 11429 at Time$1$ and Time$2$. Similar layout as Fig.~\ref{fig:decay_11158}. 
} \label{fig:decay_11429}%
\end{center}
\end{figure*}

\clearpage

\acknowledgments{We thank our anonymous referee for his/her constructive
comments that significantly improved the manuscript. We acknowledge the use of the data from HMI and AIA instruments onboard {\it Solar Dynamics Observatory} ({\it SDO}), EIT, MDI and LASCO instruments onboard {\it Solar and Heliospheric Observatory} ({\it SOHO}), and {\it Transition Region and Coronal Explorer} ({\it TRACE}). 
This work is supported by the grants from NSFC (41131065, 41574165, 41421063, 41274173, 41474151) CAS (Key Research Program KZZD-EW-01-4), MOEC (20113402110001) and the fundamental research funds for the central universities.}

\section*{APPENDIX}
\begin{appendix}
 \section{QUALITY OF NLFFF EXTRAPOLATION}\label{apd:qual}
Lorentz force ($\textbf{J}\times\textbf{B}$, where $\textbf{J}$ is the current density) and the divergence of the magnetic field ($\nabla\cdot\textbf{B}$) should 
be as small as possible to meet force-free and divergence-free condition in the NLFF coronal fields. 
We follow \citet{rliu_2016,Wheatland_2000}, using two parameters: $\theta$ (the angle between $\textbf{B}$ and $\textbf{J}$) and $\langle|f_i|\rangle$ (fractional flux increase), to measure the quality of the model fields: 
\begin{eqnarray}\label{eq:sigj}
\nonumber
&& \sigma_J=\left(\sum \limits_{i=1}^n\frac{|\textbf{J}\times\textbf{B}|_i}{B_i}\right)\Bigg/\sum\limits_{i=1}^nJ_i 
\\
&& \theta=\sin^{-1}\sigma_J
\end{eqnarray}
\begin{equation}\label{eq:fl}
\langle|f_i|\rangle=\frac{1}{n}\sum\limits_{i=1}^n\frac{|\nabla \cdot\textbf{B}|_i\Delta V_i}{B_i\cdot\Delta S_i}
\end{equation}
n is the number of the grid points, $\Delta V_i$ and $\Delta S_i$ is the volume and surface area of the $i_{th}$ cell, respective. $\sigma_J$ gives average $\sin\theta$ weighted by $J$. 
See Table.~\ref{tb:con} for $\theta$ and $\langle|f_i|\rangle$ in the aforementioned (and aftermetioned in the next three sections) model NLFF fields, which all meet force-free and divergence-free conditions.

\begin{table*}
\begin{center}
\caption{Force-free and divergence-free parameters.}\label{tb:con}
\begin{tabular}{c|c|c}
\hline
Time & $\theta$ (degree) & $\langle|f_i|\rangle \ (\times10^{-3})$\\
\hline
2011-02-14T17:10:12 & 6.70 & 2.08\\
\hline
2011-02-14T18:10:12 & 7.19 & 2.11\\
\hline
2011-02-14T19:46:20 & 6.97 & 1.90\\
\hline
2012-03-06T23:46:14 & 5.82 & 2.91\\
\hline
2012-03-07T00:58:14 & 6.78 & 2.78\\
\hline
2012-03-07T01:10:14 & 6.02 & 2.70\\
\hline
\end{tabular}
\end{center}
\end{table*}

\section{{\jkt CHANGE OF MAGNETIC PARAMETERS} DURING CME2 IN AR 11158}\label{apd:11158_cme2}

In Sec.~\ref{cases:11158}, the magnetic parameters at the source location (L1) are studied in pre-CME1 (at Time1) and post-CME1 but pre-CME2 (at Time2) corona. In this section, we perform a similar analysis in a plane perpendicular to the flux rope axis along PIL1 in the post-CME2 corona (2011-02-14T19:46:20 UT, defined as Time3), as shown in Fig.~\ref{fig:acme2_11158} (d), (e), and (f) ($Q$, $T_w$ and $\vec{B}_\parallel$, respective). The parameters at Time2 are shown in Fig.~\ref{fig:acme2_11158} (a) - (c) for comparison. The triangles mark the peak $T_w$ position, i.e., the position where the flux rope axis threading the plane. At Time3, the pronounced $Q$ boundary, strong $T_w$ region and the rotational structure around the peak $T_w$ point in the in-plane vector fields evidence a flux rope. However, the vertical magnetic flux from the strong $T_w$ region ($|T_w|\gtrsim1.25$) calculated by Equ.~\ref{eq:flux} is reduced by $77\%$ after CME2 (from $3.10\times 10^{19} $\,Mx at Time2 to $0.70\times 10^{19} $\,Mx at Time3, shown at the header of Fig.~\ref{fig:acme2_11158}(a) and (d)).  The magnetic free energy still shows a slight increase of $5.5\%$ (from $2.17 \times 10^{32} $\,erg at Time2 to $2.29 \times 10^{32} $\,erg Time3, as shown at the header of Fig.~\ref{fig:acme2_11158}(b) and (e)), which is below the uncertainty. {\jkt CME2 that is confirmed to be correlated to the source location based on observation, and decrease of twist of the rope, all evidence that the flux rope is involved into the eruption. However, the information is not enough for distinguishing whether the flux rope at Time3 is a remnant of the previous flux rope which may undergo a partial eruption accompanied by topology reconfiguration during CME2, or is a newly emerged/reformed one after CME2. Study of the CME2's eruption detail is beyond this paper's scope. 
}

\section{{\jkt CHANGE OF MAGNETIC PARAMETERS} DURING CME1 IN AR 11429}\label{apd:11429_cme1}

In Sec.~\ref{cases:11429}, the magnetic parameters at the source location of CME2 (L2) are studied in pre-CME1 (at Time1) and post-CME1 but pre-CME2 (at Time2) corona to see the possible influence from CME1 to CME2. In this section, we perform a similar analysis at the source location of CME1 (L1) to see what happened during CME1. $Q$, $T_w$ and $\vec{B}_\parallel$ are calculated in a plane perpendicular to the flux rope axis along PIL1 at Time1 (Fig.~\ref{fig:acme1_11429} (a) - (c)) and Time2 (Fig.~\ref{fig:acme1_11429} (d) - (f)), respective. Flux rope is found at PIL1 both before CME1 and after CME1. The vertical magnetic flux from the strong $T_w$ region, with a threshold of $1.25$ turns ($|T_w|\gtrsim 1.25$) , shows no significant change. However, when changing the threshold to $1.6$ turns ($|T_w|\gtrsim 1.6$), the vertical magnetic flux shows a significant reduction of $47\%$ (from $3.23\times 10^{19} $\,Mx at Time1 to $1.70\times 10^{19} $\,Mx at Time2, shown at the header of Fig.~\ref{fig:acme1_11429}(a) and (c)). {\jkt CME1 has been confirmed to be related to the source location L1 based on observation, as discussed in Sec.~\ref{subsec:d_example}, thus, the flux rope should be responsible to the eruption. It's representative axial flux with $|T_w|\gtrsim 1.6$ decreased, at the mean time, the flux with $|T_w|\gtrsim 1.25$ almost kept constant. Partial expulsion of the flux rope, accompanied by replenishment of twist through shear motion or reconnection, can explain the phenomenon. 
}

\section{{\jkt CHANGE OF MAGNETIC PARAMETERS} DURING CME2 IN AR 11429}\label{apd:11429_cme2}

 In this section, we perform a similar analysis as in Sec.~\ref{cases:11429} in a plane perpendicular to the flux rope axis along PIL2 in the post-CME2 corona (2012-03-07T01:10:12 UT, defined as Time3), as shown in Fig.~\ref{fig:acme2_11429} (d), (e), and (f) ($Q$, $T_w$ and $\vec{B}_\parallel$, respective), to see the eruption detail during CME2. The parameters at Time2 are shown in Fig.~\ref{fig:acme2_11429} (a) - (c) for comparison. 
After CME2, {\jkt there still existed a flux rope along PIL2, showing a significant topology change compared to that at Time2.}  
The vertical magnetic flux from the strong $T_w$ region ($|T_w|\gtrsim1.25$) calculated by Equ.~\ref{eq:flux} decreased by $38\%$ after CME2 (from $5.66\times 10^{19} $\,Mx at Time2 to $3.51\times 10^{19} $\,Mx at Time3, shown at the header of Fig.~\ref{fig:acme2_11429}(a) and (d)).  The magnetic free energy also shows a slight decrease of $1.5\%$ (from $8.01 \times 10^{32} $\,erg at Time2 to $7.89 \times 10^{32} $\,erg at Time3, as shown at the header of Fig.~\ref{fig:acme2_11429}(b) and (e)), which is far below the uncertainty. The flux ropes traced by the model method in our cases, and two eruptive events in \citet{rliu_2016} all show {\jkt twist remnant after the eruption. We come up two possible explanation: it is due to a partial expulsion process, or quick replenishment of twist through emergence/reformation after the eruption. The phenomenon is worth to be studied in the future.} 

\begin{figure*}
\begin{center}
\includegraphics[width=0.85\hsize]{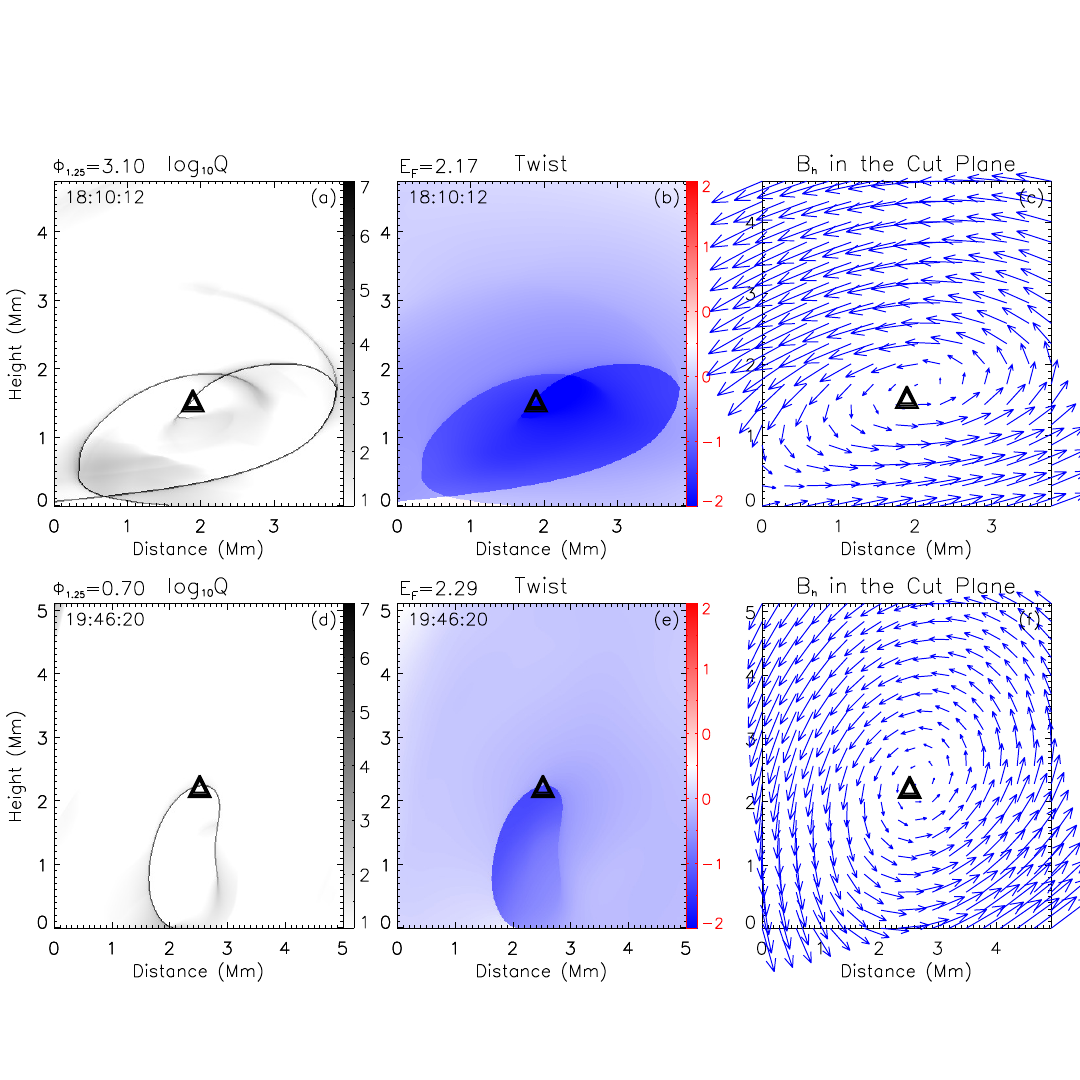}
\caption{Magnetic features in the vertical cuts, that perpendicular to the flux rope at PIL$1$ in AR 11158, at Time$2$ (post-CME1 but pre-CME2) and Time$3$ (post-CME2). Same layout as the panels (a) - (f) in Fig.~\ref{fig:rope_11158}. $\Phi_{1.25}$, given at the headers of panel (a) and (d), are vertical magnetic flux (in unit of $10^{19} $\,Mx) from the strong $T_w$ region ($|T_w|\gtrsim1.25$) at each time, respective. $E_F$, given at the header of panel (b) and (e), are magnetic free energy (in unit of $10^{32} $\,erg) at the two times.}  \label{fig:acme2_11158}
\end{center}
\end{figure*}

\begin{figure*}
\begin{center}
\includegraphics[width=0.85\hsize]{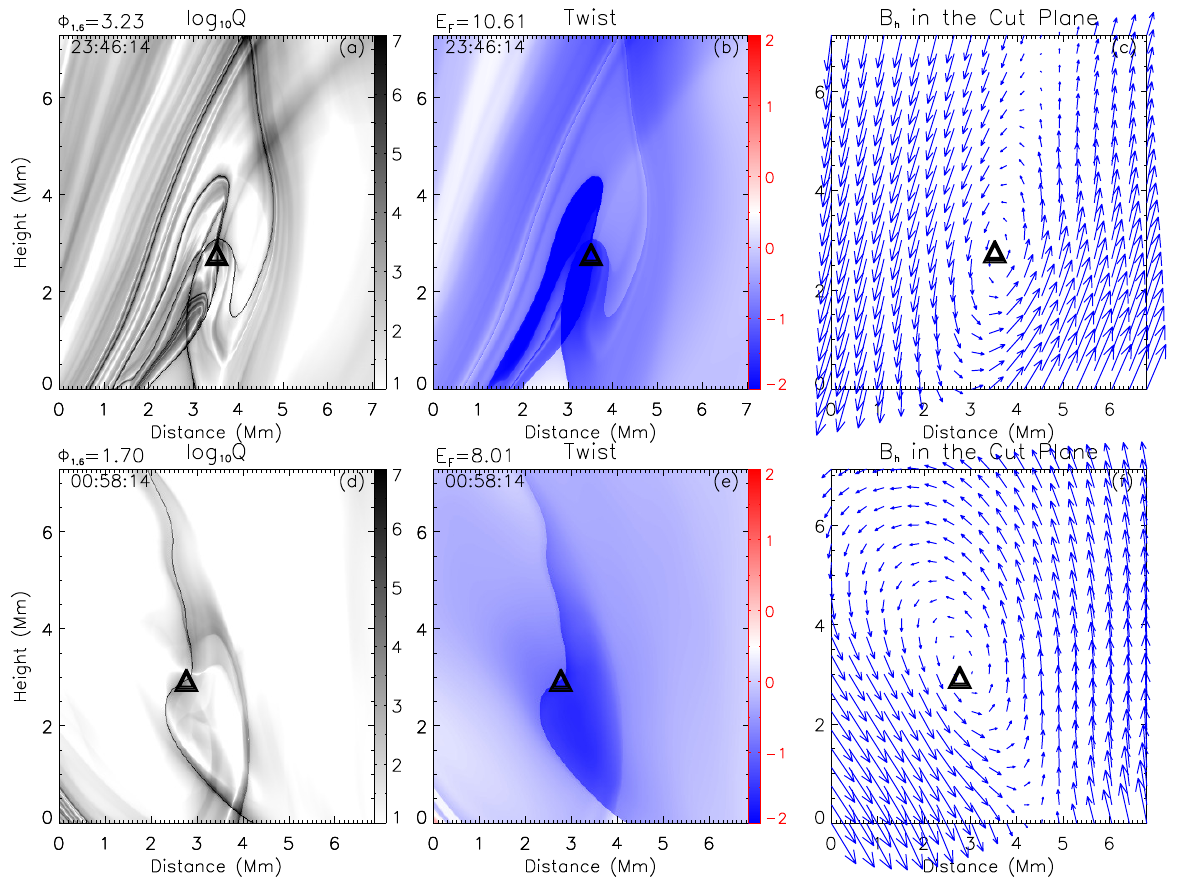}
\caption{Magnetic features in the vertical cuts, that perpendicular to the flux rope at PIL$1$ in AR 11429, at Time$1$ (pre-CME1) and Time$2$ (post-CME1 but pre-CME2). Same layout as in Fig.~\ref{fig:acme2_11158}. } \label{fig:acme1_11429}
\end{center}
\end{figure*}

\begin{figure*}
\begin{center}
\includegraphics[width=0.85\hsize]{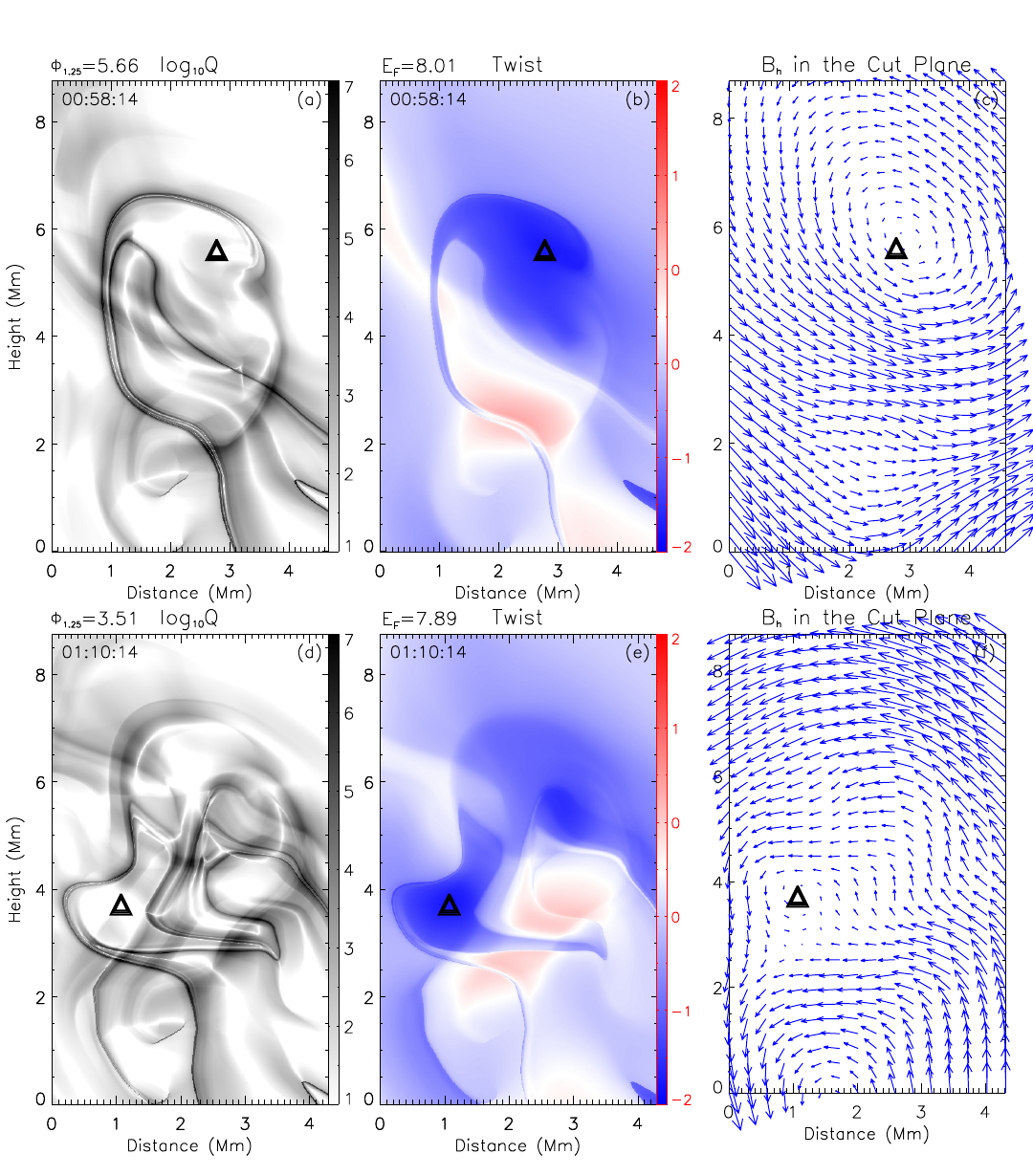}
\caption{Magnetic features in the vertical cuts, that perpendicular to the flux rope at PIL$2$ in AR 11429, at Time$2$ (post-CME1 but pre-CME2) and Time$3$ (post-CME2). Same layout as in Fig.~\ref{fig:acme2_11158}. } \label{fig:acme2_11429}
\end{center}
\end{figure*}

\end{appendix}

\clearpage
\bibliographystyle{aasjournal}
\bibliography{QC_bib3}

\begin{thebibliography}{}
\expandafter\ifx\csname natexlab\endcsname\relax\def\natexlab#1{#1}\fi
\providecommand{\url}[1]{\href{#1}{#1}}

\bibitem[{Akiyama {et~al.}(2007)Akiyama, Yashiro, \& Gopalswamy}]{Akiyama_2007}
Akiyama, S., Yashiro, S., \& Gopalswamy, N. 2007, Adv. Sp. Res., 39, 1467

\bibitem[{Amari {et~al.}(1999)Amari, Luciani, Mikic, \&
  Linker}]{Tamari_etal_1999}
Amari, T., Luciani, J.~F., Mikic, Z., \& Linker, J. 1999, Astrophys. J., 518,
  L57

\bibitem[{Balasubramaniam {et~al.}(2011)Balasubramaniam, Pevtsov, Cliver,
  Martin, \& Panasenco}]{Balasubramaniam_2011}
Balasubramaniam, K.~S., Pevtsov, A.~A., Cliver, E.~W., Martin, S.~F., \&
  Panasenco, O. 2011, Astrophys. J., 743, 202

\bibitem[{Berger \& Prior(2006)}]{MBerger_2006}
Berger, M.~A., \& Prior, C. 2006, J. Phys. A. Math. Gen., 39, 8321

\bibitem[{Brueckner {et~al.}(1995)Brueckner, Howard, Koomen, Korendyke,
  Michels, Moses, Socker, Dere, Lamy, Llebaria, Bout, Schwenn, Simnett,
  Bedford, \& Eyles}]{Brueckner_1995}
Brueckner, G.~E., Howard, R.~a., Koomen, M.~J., {et~al.} 1995, Sol. Phys., 162,
  357

\bibitem[{Chatterjee \& Fan(2013)}]{Chatterjee_2013}
Chatterjee, P., \& Fan, Y. 2013, Astrophys. J., 778, L8

\bibitem[{Chen {et~al.}(2011)Chen, Wang, Shen, Ye, Zhang, \&
  Wang}]{Chen_Wang_2011}
Chen, C., Wang, Y., Shen, C., {et~al.} 2011, J. Geophys. Res. Sp. Phys., 116,
  n/a

\bibitem[{Cheng {et~al.}(2013)Cheng, Zhang, Ding, Olmedo, Sun, Guo, \&
  Liu}]{Chengx_2013}
Cheng, X., Zhang, J., Ding, M.~D., {et~al.} 2013, Astrophys. J., 769, L25

\bibitem[{Chertok {et~al.}(2004)Chertok, Grechnev, Hudson, \&
  Nitta}]{Chertok_2004}
Chertok, I.~M., Grechnev, V.~V., Hudson, H.~S., \& Nitta, N.~V. 2004, J.
  Geophys. Res. Sp. Phys., 109, 1

\bibitem[{Chintzoglou {et~al.}(2015)Chintzoglou, Patsourakos, \&
  Vourlidas}]{Chintzoglou_2015}
Chintzoglou, G., Patsourakos, S., \& Vourlidas, A. 2015, Astrophys. J., 809, 34

\bibitem[{DeVore \& Antiochos(2008)}]{Devore_2008}
DeVore, C.~R., \& Antiochos, S.~K. 2008, Astrophys. J., 680, 740

\bibitem[{Ding {et~al.}(2013)Ding, Jiang, Zhao, \& Li}]{Ding_2013}
Ding, L., Jiang, Y., Zhao, L., \& Li, G. 2013, Astrophys. J., 763, 30

\bibitem[{Ding {et~al.}(2014)Ding, Li, Dong, Jiang, Jian, \& Gu}]{Ding_2014}
Ding, L., Li, G., Dong, L.-H., {et~al.} 2014, J. Geophys. Res. Sp. Phys., 119,
  1463

\bibitem[{Filippov(2013)}]{Filippov_2013}
Filippov, B. 2013, Astrophys. J., 773, 10

\bibitem[{Gary \& Moore(2004)}]{Gary_2004}
Gary, G.~A., \& Moore, R.~L. 2004, Astrophys. J., 611, 545

\bibitem[{Gibson \& Fan(2006)}]{Gibson_2006}
Gibson, S.~E., \& Fan, Y. 2006, Astrophys. J., 637, L65

\bibitem[{Handy {et~al.}(1999)Handy, Acton, Kankelborg, Wolfson, Akin, Bruner,
  Caravalho, Catura, Chevalier, Duncan, Edwards, Feinstein, Freeland,
  Friedlaender, Hoffmann, Hurlburt, Jurcevich, Katz, Kelly, Lemen, Levay,
  Lindgren, Mathur, Meyer, Morrison, Morrison, Nightingale, Pope, Rehse,
  Schrijver, Shine, Shing, Strong, Tarbell, Title, Torgerson, Golub,
  Bookbinder, Caldwell, Cheimets, Davis, Deluca, McMullen, Warren, Amato,
  Fisher, Maldonado, \& Parkinson}]{1999SoPh..187..229H}
Handy, B., Acton, L., Kankelborg, C., {et~al.} 1999, Sol. Phys., 187, 229

\bibitem[{Hoeksema {et~al.}(2014)Hoeksema, Liu, Hayashi, Sun, Schou, Couvidat,
  Norton, Bobra, Centeno, Leka, Barnes, \& Turmon}]{Hoeksema_etal_2014}
Hoeksema, J.~T., Liu, Y., Hayashi, K., {et~al.} 2014, Sol. Phys., 289, 3483

\bibitem[{Hood \& Priest(1981)}]{Hood_1981}
Hood, A.~W., \& Priest, E.~R. 1981, Geophys. Astrophys. Fluid Dyn., 17, 297

\bibitem[{Jing {et~al.}(2010)Jing, Tan, Yuan, Wang, Wiegelmann, Xu, \&
  Wang}]{2010ApJ...713..440J}
Jing, J., Tan, C., Yuan, Y., {et~al.} 2010, Astrophys. J., 713, 440

\bibitem[{Kaiser {et~al.}(2008)Kaiser, Kucera, Davila, {St. Cyr}, Guhathakurta,
  \& Christian}]{kaiser_etal_2008}
Kaiser, M.~L., Kucera, T.~a., Davila, J.~M., {et~al.} 2008, Space Sci. Rev.,
  136, 5

\bibitem[{Kienreich {et~al.}(2011)Kienreich, Veronig, Muhr, Temmer,
  Vr{\v{s}}nak, \& Nitta}]{Kienreich_2011}
Kienreich, I.~W., Veronig, A.~M., Muhr, N., {et~al.} 2011, Astrophys. J., 727,
  L43

\bibitem[{Kliem \& T{\"{o}}r{\"{o}}k(2006)}]{Kliem_2006}
Kliem, B., \& T{\"{o}}r{\"{o}}k, T. 2006, Phys. Rev. Lett., 96, 255002

\bibitem[{Kliem {et~al.}(2014)Kliem, T{\"{o}}r{\"{o}}k, Titov, Lionello,
  Linker, Liu, Liu, \& Wang}]{2014ApJ...792..107K}
Kliem, B., T{\"{o}}r{\"{o}}k, T., Titov, V.~S., {et~al.} 2014, Astrophys. J.,
  792, 107

\bibitem[{Lemen {et~al.}(2012)Lemen, Title, Akin, Boerner, Chou, Drake, Duncan,
  Edwards, Friedlaender, Heyman, Hurlburt, Katz, Kushner, Levay, Lindgren,
  Mathur, McFeaters, Mitchell, Rehse, Schrijver, Springer, Stern, Tarbell,
  Wuelser, Wolfson, Yanari, Bookbinder, Cheimets, Caldwell, Deluca, Gates,
  Golub, Park, Podgorski, Bush, Scherrer, Gummin, Smith, Auker, Jerram, Pool,
  Soufli, Windt, Beardsley, Clapp, Lang, \& Waltham}]{Lemen_etal_2012}
Lemen, J.~R., Title, A.~M., Akin, D.~J., {et~al.} 2012, Sol. Phys., 275, 17

\bibitem[{Li \& Zhang(2013)}]{Li_2013}
Li, T., \& Zhang, J. 2013, Astrophys. J., 778, L29

\bibitem[{Liu {et~al.}(2009)Liu, Lee, Karlick{\'{y}}, Choudhary, Deng, \&
  Wang}]{Liuc_2009}
Liu, C., Lee, J., Karlick{\'{y}}, M., {et~al.} 2009, Astrophys. J., 703, 757

\bibitem[{Liu {et~al.}(2016{\natexlab{a}})Liu, Wang, Wang, Shen, Ye, Liu, Chen,
  Zhang, \& Wang}]{Lliu_2016}
Liu, L., Wang, Y., Wang, J., {et~al.} 2016{\natexlab{a}}, Astrophys. J., 826,
  119

\bibitem[{Liu {et~al.}(2012)Liu, Kliem, T{\"{o}}r{\"{o}}k, Liu, Titov,
  Lionello, Linker, \& Wang}]{liu_2012a}
Liu, R., Kliem, B., T{\"{o}}r{\"{o}}k, T., {et~al.} 2012, Astrophys. J., 756,
  59

\bibitem[{Liu {et~al.}(2016{\natexlab{b}})Liu, Kliem, Titov, Chen, Wang, Wang,
  Liu, Xu, \& Wiegelmann}]{rliu_2016}
Liu, R., Kliem, B., Titov, V.~S., {et~al.} 2016{\natexlab{b}}, Astrophys. J.,
  818, 148

\bibitem[{Liu(2008)}]{Liuy_2008}
Liu, Y. 2008, Astrophys. J., 679, L151

\bibitem[{Lynch \& Edmondson(2013)}]{Lynch_2013}
Lynch, B.~J., \& Edmondson, J.~K. 2013, Astrophys. J., 764, 87

\bibitem[{MacTaggart \& Hood(2009)}]{Mactaggart_2009}
MacTaggart, D., \& Hood, A.~W. 2009, Astron. Astrophys., 508, 445

\bibitem[{Moon {et~al.}(2003{\natexlab{a}})Moon, Chae, Wang, \&
  Park}]{Moon_2003b}
Moon, Y.-J., Chae, J., Wang, H., \& Park, Y. 2003{\natexlab{a}}, Adv. Sp. Res.,
  32, 1953

\bibitem[{Moon {et~al.}(2003{\natexlab{b}})Moon, Choe, Wang, \&
  Park}]{Moon_2003}
Moon, Y.-J., Choe, G.~S., Wang, H., \& Park, Y.~D. 2003{\natexlab{b}},
  Astrophys. J., 588, 1176

\bibitem[{Nitta \& Hudson(2001)}]{Nitta_2001}
Nitta, N.~V., \& Hudson, H.~S. 2001, Geophys. Res. Lett., 28, 3801

\bibitem[{Pesnell {et~al.}(2012)Pesnell, Thompson, \&
  Chamberlin}]{Pesnell_2012}
Pesnell, W.~D., Thompson, B.~J., \& Chamberlin, P.~C. 2012, Sol. Phys., 275, 3

\bibitem[{Schmieder(2006)}]{Schmieder_2006}
Schmieder, B. 2006, J. Astrophys. Astron., 27, 139

\bibitem[{Schou {et~al.}(2012)Schou, Scherrer, Bush, Wachter, Couvidat,
  Rabello-Soares, Bogart, Hoeksema, Liu, Duvall, Akin, Allard, Miles, Rairden,
  Shine, Tarbell, Title, Wolfson, Elmore, Norton, \& Tomczyk}]{Schou_2012}
Schou, J., Scherrer, P.~H., Bush, R.~I., {et~al.} 2012, Sol. Phys., 275, 229

\bibitem[{Schrijver(2009)}]{Schrijver_2009}
Schrijver, C.~J. 2009, Adv. Sp. Res., 43, 739

\bibitem[{Schrijver \& Title(2011)}]{Schrijver_2011}
Schrijver, C.~J., \& Title, A.~M. 2011, J. Geophys. Res. Sp. Phys., 116, n/a

\bibitem[{Shen {et~al.}(2013)Shen, Li, Kong, Hu, Sun, Ding, Chen, Wang, \&
  Xia}]{Shenc_2013}
Shen, C., Li, G., Kong, X., {et~al.} 2013, Astrophys. J., 763, 114

\bibitem[{Soenen {et~al.}(2009)Soenen, Zuccarello, Jacobs, Poedts, Keppens, \&
  van~der Holst}]{Soenen_2009}
Soenen, A., Zuccarello, F.~P., Jacobs, C., {et~al.} 2009, Astron. Astrophys.,
  501, 1123

\bibitem[{Su {et~al.}(2014)Su, Jing, Wang, Wiegelmann, \&
  Wang}]{2014ApJ...788..150S}
Su, J.~T., Jing, J., Wang, S., Wiegelmann, T., \& Wang, H.~M. 2014, Astrophys.
  J., 788, 150

\bibitem[{Su {et~al.}(2015)Su, van Ballegooijen, McCauley, Ji, Reeves, \&
  DeLuca}]{SuY_2015}
Su, Y., van Ballegooijen, A., McCauley, P., {et~al.} 2015, Astrophys. J., 807,
  144

\bibitem[{Subramanian \& Dere(2001)}]{SUBRAMANIAN_2001}
Subramanian, P., \& Dere, K.~P. 2001, Astrophys. J., 561, 372

\bibitem[{Sun {et~al.}(2012)Sun, Hoeksema, Liu, Chen, \& Hayashi}]{Sun_2012a}
Sun, X., Hoeksema, J.~T., Liu, Y., Chen, Q., \& Hayashi, K. 2012, Astrophys.
  J., 757, 149

\bibitem[{Sun {et~al.}(2015)Sun, Bobra, Hoeksema, Liu, Li, Shen, Couvidat,
  Norton, \& Fisher}]{Sun_etal_2015}
Sun, X., Bobra, M.~G., Hoeksema, J.~T., {et~al.} 2015, Astrophys. J., 804, L28

\bibitem[{Thalmann {et~al.}(2015)Thalmann, Su, Temmer, \&
  Veronig}]{Thalmann_etal_2015}
Thalmann, J.~K., Su, Y., Temmer, M., \& Veronig, a.~M. 2015, Astrophys. J.,
  801, L23

\bibitem[{Thalmann {et~al.}(2008)Thalmann, Wiegelmann, \&
  Raouafi}]{Thalmann_2008a}
Thalmann, J.~K., Wiegelmann, T., \& Raouafi, N.-E. 2008, Astron. Astrophys.,
  488, L71

\bibitem[{Tian {et~al.}(2002)Tian, Liu, \& Wang}]{Tian_2002}
Tian, L., Liu, Y., \& Wang, J. 2002, Sol. Phys., 209, 361

\bibitem[{Titov(2007)}]{Titov_2007}
Titov, V.~S. 2007, Astrophys. J., 660, 863

\bibitem[{Titov \& D{\'{e}}moulin(1999)}]{Titov_1999}
Titov, V.~S., \& D{\'{e}}moulin, P. 1999, Astron. Astrophys., 351, 707

\bibitem[{Titov {et~al.}(2002)Titov, Hornig, \& D{\'{e}}moulin}]{Titov_2002}
Titov, V.~S., Hornig, G., \& D{\'{e}}moulin, P. 2002, J. Geophys. Res. Sp.
  Phys., 107, SSH 3

\bibitem[{Titov {et~al.}(1993)Titov, Priest, \& D{\'{e}}moulin}]{Titov_1993}
Titov, V.~S., Priest, E.~R., \& D{\'{e}}moulin, P. 1993, Astron. Astrophys.,
  276, 564

\bibitem[{T{\"{o}}r{\"{o}}k \& Kliem(2003)}]{Torok_2003}
T{\"{o}}r{\"{o}}k, T., \& Kliem, B. 2003, Astron. Astrophys., 406, 1043

\bibitem[{T{\"{o}}r{\"{o}}k \& Kliem(2005)}]{Torok_Kliem_2005}
T{\"{o}}r{\"{o}}k, T., \& Kliem, B. 2005, Astrophys. J., 630, L97

\bibitem[{T{\"{o}}r{\"{o}}k {et~al.}(2011)T{\"{o}}r{\"{o}}k, Panasenco, Titov,
  Miki{\'{c}}, Reeves, Velli, Linker, \& {De Toma}}]{Torok_2011}
T{\"{o}}r{\"{o}}k, T., Panasenco, O., Titov, V.~S., {et~al.} 2011, Astrophys.
  J., 739, L63

\bibitem[{Wang {et~al.}(2011)Wang, Chen, Gui, Shen, Ye, \& Wang}]{wang_2011_st}
Wang, Y., Chen, C., Gui, B., {et~al.} 2011, J. Geophys. Res. Sp. Phys., 116,
  n/a

\bibitem[{Wang {et~al.}(2013)Wang, Liu, Shen, Liu, Ye, \& Wang}]{Wang_2013}
Wang, Y., Liu, L., Shen, C., {et~al.} 2013, Astrophys. J., 763, L43

\bibitem[{Wang \& Zhang(2007)}]{Wang_zhang_2007}
Wang, Y., \& Zhang, J. 2007, Astrophys. J., 665, 1428

\bibitem[{Webb \& Howard(2012)}]{Webb_2012}
Webb, D.~F., \& Howard, T.~A. 2012, Living Rev. Sol. Phys., 9, 3

\bibitem[{Wheatland {et~al.}(2000)Wheatland, Sturrock, \&
  Roumeliotis}]{Wheatland_2000}
Wheatland, M.~S., Sturrock, P.~a., \& Roumeliotis, G. 2000, Astrophys. J., 540,
  1150

\bibitem[{Wiegelmann(2004)}]{Wiegelmann_2004}
Wiegelmann, T. 2004, Sol. Phys., 219, 87

\bibitem[{Wiegelmann {et~al.}(2012)Wiegelmann, Thalmann, Inhester, Tadesse,
  Sun, \& Hoeksema}]{Wiegelmann_2012}
Wiegelmann, T., Thalmann, J.~K., Inhester, B., {et~al.} 2012, Sol. Phys., 281,
  37

\bibitem[{Yang {et~al.}(2012)Yang, Jiang, Zheng, Bi, Hong, \& Yang}]{Yang_2012}
Yang, J., Jiang, Y., Zheng, R., {et~al.} 2012, Astrophys. J., 745, 9

\bibitem[{Yashiro(2004)}]{2004JGRA..109.7105Y}
Yashiro, S. 2004, J. Geophys. Res., 109, A07105

\bibitem[{Yashiro {et~al.}(2008)Yashiro, Michalek, Akiyama, Gopalswamy, \&
  Howard}]{2008ApJ...673.1174Y}
Yashiro, S., Michalek, G., Akiyama, S., Gopalswamy, N., \& Howard, R.~A. 2008,
  Astrophys. J., 673, 1174

\bibitem[{Zhang \& Wang(2002)}]{Zhang_Wang_2002}
Zhang, J., \& Wang, J. 2002, Astrophys. J., 566, L117

\bibitem[{Zhou {et~al.}(2003)Zhou, Wang, \& Cao}]{Zhougp_2003}
Zhou, G., Wang, J., \& Cao, Z. 2003, Astron. Astrophys., 397, 1057

\bibitem[{Zuccarello {et~al.}(2009)Zuccarello, Romano, Farnik, Karlicky,
  Contarino, Battiato, Guglielmino, Comparato, \&
  Ugarte-Urra}]{Zuccarello_2009}
Zuccarello, F., Romano, P., Farnik, F., {et~al.} 2009, Astron. Astrophys., 493,
  629

\bibitem[{Zuccarello {et~al.}(2016)Zuccarello, Aulanier, \&
  Gilchrist}]{Zuccarello_2016}
Zuccarello, F.~P., Aulanier, G., \& Gilchrist, S.~A. 2016, Astrophys. J., 821,
  L23

\end{thebibliography}

\clearpage

\end{document}